\documentclass[11pt,a4paper]{article}
\pdfoutput=1

\usepackage{jcappub}
\usepackage{graphicx}
\usepackage{xcolor}
\usepackage{mathrsfs,mathtools}
\usepackage{physics,amssymb}
\usepackage{siunitx}
\usepackage{bm}
\usepackage{braket}
\usepackage{listings}
\usepackage{cases}
\usepackage{comment}
\usepackage{soul}
\usepackage{cancel}
\usepackage{cases}
\usepackage[utf8]{inputenc}
\usepackage{url}
\usepackage{xspace}
\usepackage{acronym}
\usepackage{aas_macros}
\usepackage{tikz-feynhand}
\usepackage{subcaption}

\hypersetup{colorlinks=true
,urlcolor=DARKBLUE
,anchorcolor=DARKBLUE
,citecolor=DARKBLUE
,filecolor=DARKBLUE
,linkcolor=DARKBLUE
,menucolor=DARKBLUE
,linktocpage=true
,pdfproducer=medialab
,pdfa=true
}

\newcommand{\ee}{\mathrm{e}}
\newcommand{\Mpl}{M_\mathrm{Pl}}
\newcommand{\ns}{n_{\mathrm{s}}}

\newcommand{\fNL}{f_\mathrm{NL}}
\newcommand{\fNLs}{f_\mathrm{NLs}}
\newcommand{\fNLeta}{f_{\mathrm{NL}\eta}}
\newcommand{\fNLH}{f_{\mathrm{NL}H}}
\newcommand{\SR}{\mathrm{SR}}
\newcommand{\CR}{\mathrm{CR}}
\newcommand{\bulk}{\mathrm{bulk}}
\newcommand{\EoM}{\mathrm{EoM}}
\newcommand{\UV}{\mathrm{UV}}

\newcommand{\calA}{\mathcal{A}}
\newcommand{\uB}{\mathrm{B}}
\newcommand{\calB}{\mathcal{B}}

\newcommand{\uc}{\mathrm{c}}

\newcommand{\ue}{\mathrm{e}}

\newcommand{\uI}{\mathrm{I}}

\newcommand{\bfk}{\mathbf{k}}
\newcommand{\uL}{\mathrm{L}}
\newcommand{\calL}{\mathcal{L}}

\newcommand{\un}{\mathrm{n}}
\newcommand{\calO}{\mathcal{O}}

\newcommand{\calP}{\mathcal{P}}

\newcommand{\uS}{\mathrm{S}}
\newcommand{\us}{\mathrm{s}}
\newcommand{\bfx}{\mathbf{x}}

\newcommand{\ui}{\mathrm{i}}

\newcommand{\beae}[1]{\begin{equation}\begin{aligned} #1 \end{aligned}\end{equation}}
\newcommand{\bege}[1]{\begin{equation}\begin{gathered} #1 \end{gathered}\end{equation}}
\newcommand{\bae}[1]{\begin{align} #1 \end{align}}
\newcommand{\bce}[1]{\begin{cases} #1 \end{cases}}
\newcommand{\dps}{\displaystyle}
\newcommand{\bfe}[4]{
\begin{figure} 
	\centering
	\includegraphics[#1]{#2}
	\caption{#3}
	\label{#4}
\end{figure}}
\newcommand{\bme}[1]{\begin{multline} #1 \end{multline}}
\newcommand{\bmte}[1]{\begin{multlined}[t] #1 \end{multlined}}
\newcommand{\bmbe}[1]{\begin{multlined}[b] #1 \end{multlined}}

\definecolor{MONZA}{HTML}{CF000F}
\definecolor{DARKBLUE}{HTML}{00008b}
\definecolor{DARKMAGENTA}{HTML}{8b008b}

\acrodef{PBH}{primordial black hole}
\acrodef{GW}{gravitational wave}
\acrodef{CMB}{cosmic microwave background}
\acrodef{ADM}{Arnowitt, Deser, and Misner}
\acrodef{EoM}{equation of motion}

\newcommand{\f}[2]{\frac{#1}{#2}}  
\newcommand{\mk}[1]{\left( #1 \right)}  
\newcommand{\kk}[1]{\left[ #1 \right]}

\newcommand{\beq}{\begin{equation}}  
\newcommand{\eeq}{\end{equation}}
\newcommand{\e}{\epsilon}

\newcommand{\Ck}{\calA_k}
\newcommand{\Dk}{\calB_k}

\newcommand{\nucr}{{\nu_{\rm CR}}}

\newcommand{\crit}{\mathrm{crit}}

\newcommand{\mO}{\mathcal{O}}

\renewcommand{\Re}{{\rm Re}\,}
\renewcommand{\Im}{{\rm Im}\,}

\title{Squeezed bispectrum and one-loop corrections in transient constant-roll inflation}

\author[a]{Hayato Motohashi}
\author[b,c,d]{and Yuichiro Tada}

\affiliation[a]{Division of Liberal Arts, Kogakuin University, 2665-1 Nakano-machi, Hachioji, Tokyo, 192-0015, Japan}
\affiliation[b]{Institute for Advanced Research, Nagoya University,
Furo-cho Chikusa-ku, 
Nagoya 464-8601, Japan}
\affiliation[c]{Department of Physics, Nagoya University, 
Furo-cho Chikusa-ku,
Nagoya 464-8602, Japan}
\affiliation[d]{Theory Center, IPNS, KEK, 
1-1 Oho, Tsukuba, 
Ibaraki 305-0801, Japan}

\emailAdd{motohashi@cc.kogakuin.ac.jp}
\emailAdd{tada.yuichiro.y8@f.mail.nagoya-u.ac.jp}

\abstract{
In canonical single-field inflation, the production of primordial black holes (PBH) requires a transient violation of the slow-roll condition.  The transient ultra slow-roll inflation is an example of such scenarios, and more generally, one can consider the transient constant-roll inflation.  We investigate the squeezed bispectrum in the transient constant-roll inflation and find that Maldacena's consistency relation holds for a sufficiently long-wavelength mode, whereas it is violated for modes around the peak scale for the non-attractor case.  We also demonstrate how the one-loop corrections are modified compared to the case of the transient ultra slow-roll inflation, focusing on representative one-loop terms originating from a time derivative of the second slow-roll parameter in the cubic action.  We find that the perturbativity requirement on those terms does not rule out the production of PBH from the transient constant-roll inflation.
Therefore, it is a simple counterexample of the recently claimed no-go theorem of PBH production from single-field inflation.
}

\begin{document}
\maketitle

\section{Introduction}

\Acp{PBH}~\cite{Zeldovich:1967lct,Hawking:1971ei,Carr:1974nx,Carr:1975qj}, hypothetical black holes formed in the early universe before any star-forming, are attracting attention more and more.
They can explain a dominant component of dark matter~\cite{Carr:2020gox}, black-hole-merger events discovered by the LIGO--Virgo--KAGRA collaborations~\cite{Sasaki:2018dmp}, microlensing events towards the Galactic bulge generated by planetary-mass objects~\cite{Niikura:2019kqi},
possible origins of supermassive black holes~\cite{Serpico:2020ehh}, early massive galaxies discovered by the James Webb Space Telescope~\cite{Carr:2018rid,Liu:2022bvr}, etc. (see a recent review~\cite{Escriva:2022duf}).

One of the main formation scenarios of \acp{PBH} is the collapse of order-unity overdensities associated with large primordial perturbations generated, e.g., by cosmic inflation.
In the canonical single-field inflation scenario, the enhancement of the primordial perturbation amplitude necessary for production of sizable amount of PBHs requires a transient violation of slow-roll~\cite{Motohashi:2017kbs}
\bae{ 
	\f{\Delta \ln \e}{\Delta N} \lesssim -0.4 .
}
where $\epsilon=-\dot{H}/H^2$ is the first slow-roll parameter and $N=\int H\dd{t}$ is the e-folding number.
It implies that a large value of the second slow-roll parameter $\eta=\dot{\epsilon}/(\epsilon H)$. 
A special case $\eta=-6$ is known as the ultra slow-roll limit \cite{Tsamis:2003px,Kinney:2005vj}, and 
the transient ultra slow-roll inflation has been extensively studied recently as one of the representative models of the \acp{PBH} production~\cite{Garcia-Bellido:2017mdw,Ezquiaga:2017fvi,Germani:2017bcs,Motohashi:2017kbs,Ballesteros:2017fsr,Cicoli:2018asa,Biagetti:2018pjj,Dalianis:2018frf,Byrnes:2018txb,Passaglia:2018ixg,Bhaumik:2019tvl,Cheong:2019vzl,Ragavendra:2020sop,Figueroa:2020jkf,Pattison:2021oen,Inomata:2021uqj,Inomata:2021tpx,Geller:2022nkr}. 
More generally, the constant $\eta$ phase is called \emph{constant-roll inflation}. 
The constant-roll inflation~\cite{Martin:2012pe,Motohashi:2014ppa,Motohashi:2017vdc,Motohashi:2019tyj} allows an exact solution satisfying a condition of the constant rate of roll, $\ddot\phi/(H\dot\phi) = \beta$ with constant $\beta$, for which $\eta\approx 2\beta$ holds.  It is often used a scenario to generate a red-tilted spectrum compatible with \ac{CMB} observations~\cite{Motohashi:2014ppa,Motohashi:2017aob,GalvezGhersi:2018haa,Stein:2022cpk}, but using a different parameter range, it can also generate a blue-tilted spectrum in non-/attractor dynamics, which can be utilized to \ac{PBH} production~\cite{Motohashi:2019rhu,Mishra:2019pzq,Ozsoy:2020kat,Karam:2022nym}.

In the context of \ac{PBH} production, not only the amplitude of the power spectrum but also the non-Gaussian feature of the primordial perturbation is important as it affects the mean abundance (see, e.g., Refs.~\cite{Bullock:1996at,Ivanov:1997ia,Yokoyama:1998pt,Hidalgo:2007vk,Byrnes:2012yx,Bugaev:2013vba,Young:2015cyn,Nakama:2016gzw,Ando:2017veq,Franciolini:2018vbk,Atal:2018neu,Passaglia:2018ixg,Atal:2019cdz,Atal:2019erb,Yoo:2019pma,Taoso:2021uvl,Kitajima:2021fpq,Escriva:2022pnz}), the spatial distribution (clustering) (see, e.g., Refs.~\cite{Chisholm:2005vm,Young:2014ana,Young:2014oea,Tada:2015noa,Young:2015kda,Suyama:2019cst,Young:2019gfc}), the corresponding induced \ac{GW} (see, e.g., Refs.~\cite{Cai:2018dig,Unal:2018yaa,Yuan:2020iwf,Adshead:2021hnm,Abe:2022xur} and also Ref.~\cite{Domenech:2021ztg} for a recent review), etc.  
Also, the one-loop corrections have been actively discussed recently in the context of whether the \ac{PBH} production in the canonical single-filed inflation is ruled out from the perturbativity requirement~\cite{Kristiano:2022maq,Kristiano:2023scm,Riotto:2023hoz,Riotto:2023gpm,Choudhury:2023vuj,Choudhury:2023jlt,Choudhury:2023rks,Firouzjahi:2023aum,Firouzjahi:2023ahg}.  
While previous researches mainly focused on the \ac{PBH} production in the transient ultra slow-roll inflation, as stressed above, it is not the unique option even within the canonical single-field inflation.
The non-Gaussianity and one-loop corrections in other single-field scenarios remain unclear.

In this paper, we investigate the non-Gaussianity and one-loop corrections in the transient constant-roll model.  After reviewing the transient constant-roll inflation in \S\ref{sec:conroll}, we consider the squeezed bispectrum in \S\ref{sec:bisp}, evaluated on both regimes after and during the constant roll phase.  We clarify the cases where the Maldacena's consistency relation~\cite{Maldacena:2002vr} is satisfied or violated. In \S\ref{sec:loop}, focusing on representative terms, which originate from $\dot\eta$ term in the cubic action, we demonstrate how the one-loop corrections are modified compared to the case of the transient ultra slow-roll inflation.  We find that the perturbativity requirement on those terms does not rule out the production of \acp{PBH} from the transient constant-roll inflation.  \S\ref{sec:conc} is devoted to conclusions.
Throughout the paper, we work in the Planck units where $c=\hbar=\Mpl=1$.

\section{Transient constant-roll inflation}
\label{sec:conroll}

Let us review brief features of the constant-roll phase, assuming a toy transient model.
In this model, the constant-roll phase is inserted in the slow-roll inflation and each phase is characterized by the second slow-roll parameter
\bae{
	\eta=\frac{\dot{\epsilon}}{\epsilon H}=\bce{
		0, & \tau<\tau_\us, \\
		2\beta, & \tau_\us\leq\tau<\tau_\ue, \\
		0, & \tau_\ue\leq\tau,
	}
}
where $\beta$ is a constant.\footnote{
The realisation of the sharp transitions between the slow-roll and contant-roll regimes is shown in Ref.~\cite{Motohashi:2019rhu}.  One can also reconstruct the exact step-function-like transitions in the Hamilton--Jacobi approach (see Appendix~\ref{sec: appendix}).}
To generate a blue-tilted spectrum for \ac{PBH} production, we focus on the constant-roll model with a negative value of $\beta$.
$\tau<0$ is the conformal time and $\tau_{\us(\ue)}$ represents the onset (end) of the constant-roll phase.
$\beta=-3$ corresponds to the ultra slow-roll inflation. While $\beta<-3/2$ yields a non-attractor dynamics, $-3/2<\beta<0$ follows an attractor solution and the curvature perturbation gets frozen in the superhorizon limit~\cite{Motohashi:2014ppa,Morse:2018kda,Gao:2019sbz,Lin:2019fcz,Motohashi:2019rhu}.
We assume that the first slow-roll parameter $\epsilon$ is small enough and the Hubble parameter $H=\dot{a}/a$ ($a$ is the global scale factor) is almost constant during inflation. 
It is then solved as
\bae{
	\epsilon=-\frac{\dot{H}}{H^2}=\bce{
		\epsilon_{\SR1}, & \tau<\tau_\us, \\
		\epsilon_{\SR1}\pqty{\frac{\tau}{\tau_\us}}^{-2\beta}, & \tau_\us\leq\tau<\tau_\ue, \\
		\epsilon_{\SR2}=\epsilon_{\SR1}\pqty{\frac{\tau_\ue}{\tau_\us}}^{-2\beta}, & \tau_\ue\leq\tau,
	}
}
with the constant initial value $\epsilon_{\SR1}$.
The slow-roll parameter $\epsilon$ decreases during the constant-roll phase and then the power spectrum of the curvature perturbation can be enhanced on the corresponding scales.

Let us see the linear dynamics of the curvature perturbation on the comoving slice, $\zeta(\tau,\bfx)$.
The corresponding quantum operator $\hat{\zeta}_\uI(\tau,\bfx)$ (i.e., $\hat{\zeta}$ in the interaction picture) is expanded by the annihilation-creation operator $\hat{a}_\bfk/\hat{a}_\bfk^\dagger$ as
\bae{
	\hat{\zeta}_{\uI,\bfk}(\tau)=\int\frac{\dd[3]\bfk}{(2\pi)^3}\pqty{\zeta_k(\tau)\hat{a}_\bfk+\zeta_k^*(\tau)\hat{a}_{-\bfk}^\dagger}\ee^{i\bfk\cdot\bfx}.
}
The annihilation-creation operator satisfies the commutation relation
\bae{
	[\hat{a}_\bfk,\hat{a}_{\bfk^\prime}^\dagger]=(2\pi)^3\delta^{(3)}(\bfk-\bfk^\prime) \qc
	[\hat{a}_\bfk,\hat{a}_{\bfk^\prime}]=[\hat{a}^\dagger_\bfk,\hat{a}^\dagger_{\bfk^\prime}]=0.
}
The mode function $\zeta_k(\tau)$ is governed by the Mukhanov--Sasaki equation
\bae{
	v_k^{\prime\prime}+\pqty{k^2-\frac{z^{\prime\prime}}{z}}v_k=0,
}
where the Mukhanov--Sasaki variable $v_k$ is related to $\zeta_k$ by $v_k=z\zeta_k$ with $z=a\sqrt{2\epsilon}$.
The prime denotes the conformal time derivative.
The effective mass term $z^{\prime\prime}/z$ is expressed in terms of slow-roll parameters as
\bae{
	\f{z''}{z}=a^2H^2\mk{2-\e_1+\f{3}{2}\e_2+\f{1}{4}\e_2^2-\f{1}{2}\e_1\e_2+\f{1}{2}\e_2\e_3},
}
where $\e_1=\e$ and $\e_{n+1}=H^{-1}\dd{\ln \e_n}/\dd{t}$ (i.e., $\epsilon_2=\eta$).
Both in the slow-roll (all slow-roll parameters are negligible) and constant-roll phases (slow-roll parameters except for $\eta$ are negligible), this can be rewritten as
\bae{
	\frac{z^{\prime\prime}}{z}=\frac{\nu^2-1/4}{\tau^2},
}
with
\bae{
    \nu=\bce{
        \nu_\SR=3/2, & \text{slow-roll phase,} \\
        \nu_\CR=\abs{3/2+\beta}, & \text{constant-roll phase},
    }
}
The solution of the Mukhanov--Sasaki equation 
is given by a superposition of the Hankel functions $\sqrt{-k\tau}H_\nu^{(1)}(-k\tau)$ and $\sqrt{-k\tau}H_\nu^{(2)}(-k\tau)$. 
Particularly in the slow-roll phase, the solution is simplified as
\bae{
    \zeta_k=\frac{iH}{2\sqrt{\epsilon k^3}}\bqty{\calA_k\ee^{-ik\tau}(1+ik\tau)-\calB_k\ee^{ik\tau}(1-ik\tau)},
}
with the constants of integration, $\calA_k$ and $\calB_k$.
These coefficients are fixed by the Bunch--Davies initial condition
\bae{\label{eq: Bunch Davies}
    v_k\underset{\tau\to-\infty}{\to}\frac{\ee^{-ik\tau}}{\sqrt{2k}},
}
and junction conditions at $\tau_\us$ and $\tau_\ue$,
\bae{\label{eq: junction}
    \zeta_k(\tau_{\us(\ue)}-0)=\zeta_k(\tau_{\us(\ue)}+0) \qc
    \zeta_k^\prime(\tau_{\us(\ue)}-0)=\zeta_k^\prime(\tau_{\us(\ue)}+0).
}

The (tree-level) power spectrum of the curvature perturbation is defined by
\bae{
    P_\zeta(k)=\abs{\zeta_k}^2.
}
The following dimensionless power is also useful.
\bae{\label{eq: calPz}
    \calP_\zeta(k)=\frac{k^3}{2\pi^2}P_\zeta(k).
}
We show example spectra for various values of $\beta$ in Fig.~\ref{fig: calPCR}.
One can confirm that though it is small enough as $\calP_\zeta\sim2\times10^{-9}$ to be consistent with the \ac{CMB} observation on large scales $\gtrsim\SI{1}{Mpc}$~\cite{Planck:2018jri}, the power spectrum can be sufficiently large for \ac{PBH} formation on a small scale even in the attractor models $-3/2<\beta<0$.

\bfe{width=0.7\hsize}{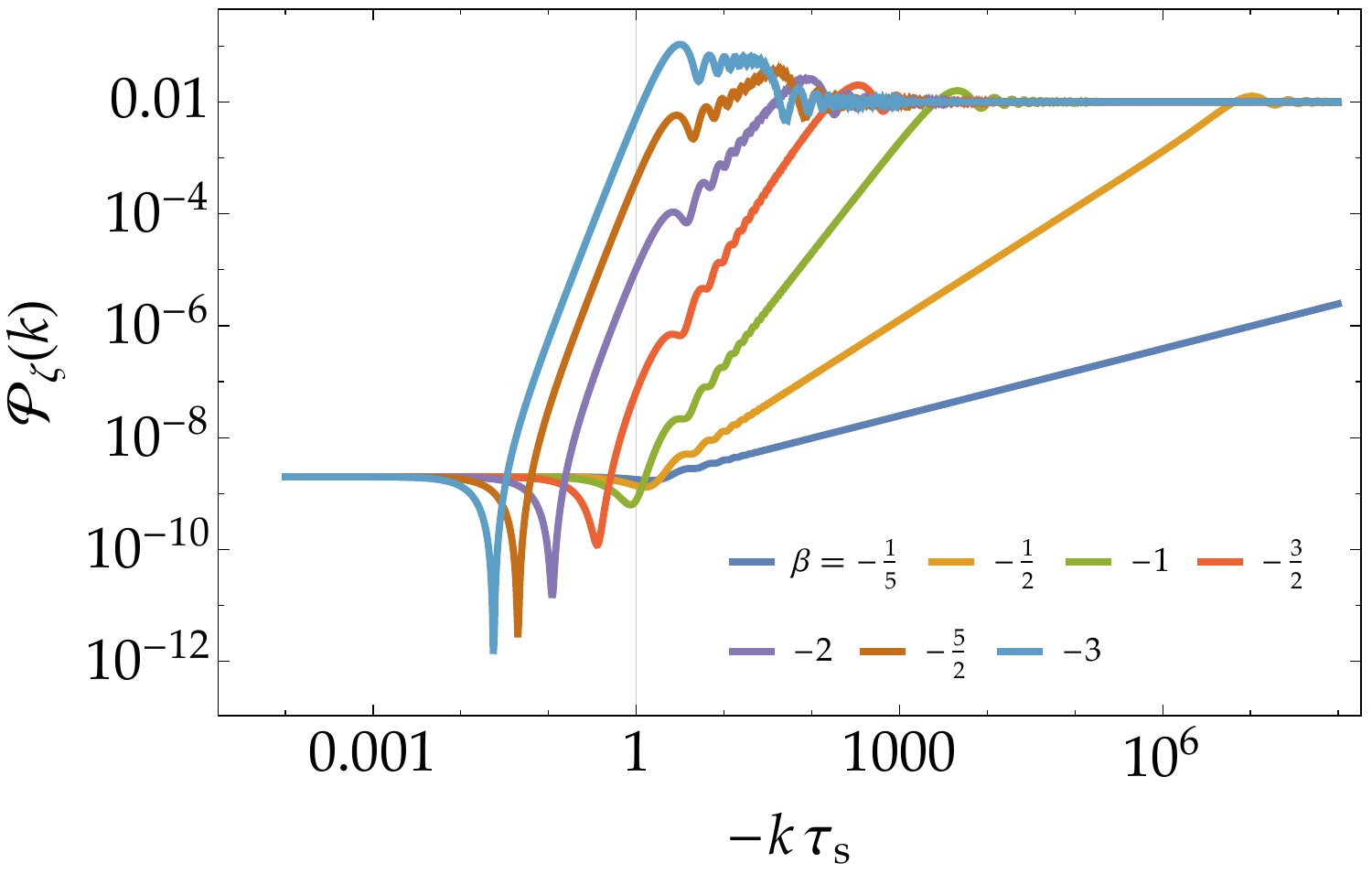}{Example power spectra~\eqref{eq: calPz} in the late-time limit $\tau\to0$ for $\beta=-1/5$ (blue), $-1/2$ (yellow), $-1$ (green), $-3/2$ (red), $-2$ (purple), $-5/2$ (brown), and $-3$ (turquoise). $\tau_\ue/\tau_\us$ is determined by the condition $\epsilon_{\SR2}/\epsilon_{\SR1}=\pqty{\tau_\ue/\tau_\us}^{-2\beta}=2\times10^{-9}/10^{-2}$ so that $\calP_\zeta\sim2\times10^{-9}$ on a large scale and $\calP_\zeta\sim10^{-2}$ on a small scale.}{fig: calPCR}

\section{Squeezed bispectrum}
\label{sec:bisp}

The bispectrum $B_\zeta(k_1,k_2,k_3)$, a three-point function in Fourier space, is defined by
\bae{
    \braket{\hat{\zeta}_{\bfk_1}\hat{\zeta}_{\bfk_2}\hat{\zeta}_{\bfk_3}}=(2\pi)^3\delta^{(3)}(\bfk_1+\bfk_2+\bfk_3)B_\zeta(k_1,k_2,k_3),
}
where $\hat{\zeta}_\bfk$ is the Fourier mode of $\hat{\zeta}(\bfx)$.
In particular, its squeezed limit where one momentum is much smaller than the others is a useful indicator of the physics sourcing the curvature perturbation.
It also affects the \ac{PBH} formation.
For example, the following local-type non-Gaussianity is known as a phenomenological model showing a non-zero squeezed bispectrum:
\bae{\label{eq: local fNL}
    \hat{\zeta}(\bfx)\approx\hat{g}(\bfx)+\frac{3}{5}\fNL\hat{g}^2(\bfx),
}
where $\hat{g}$ is a Gaussian random field and the coefficient $\fNL$ is called the non-linearity parameter.
At the leading order in the expansion in $\hat{g}$, the bispectrum and the non-linearity parameter are related by
\bae{
    B_\zeta(k_1,k_2,k_3)\underset{k_1/k_2\ll1}{\to}\frac{12}{5}\fNL P_\zeta(k_1)P_\zeta(k_2).
}
For a small-enough perturbation $\hat{g}\ll1$, the non-Gaussian part (the second term) in Eq.~\eqref{eq: local fNL} is subdominant as long as the non-linearity parameter is not so large, $\fNL\lesssim\calO(1)$.
However, because \acp{PBH} are caused by the order-unity curvature perturbation, even the not-so-large non-Gaussianity $\fNL\sim1$ can significantly affect the \ac{PBH} abundance (see, e.g., Refs.~\cite{Bullock:1996at,Ivanov:1997ia,Yokoyama:1998pt,Hidalgo:2007vk,Byrnes:2012yx,Bugaev:2013vba,Young:2015cyn,Nakama:2016gzw,Ando:2017veq,Franciolini:2018vbk,Cai:2018dig,Atal:2018neu,Passaglia:2018ixg,Atal:2019cdz,Atal:2019erb,Yoo:2019pma,Taoso:2021uvl,Kitajima:2021fpq,Escriva:2022pnz,Abe:2022xur}).
The squeezed bispectrum also means the correlation between the long- and short-wavelength modes.
Hence it can generate the large-scale modulation of the \ac{PBH} spatial distribution known as the \ac{PBH} bias or clustering (see, e.g., Refs.~\cite{Chisholm:2005vm,Young:2014ana,Young:2014oea,Tada:2015noa,Young:2015kda,Suyama:2019cst,Young:2019gfc}).
From this perspective, we investigate the details of the squeezed bispectrum in the transient constant-roll inflation model in this section.
We often use the generalized non-linearity parameter defined by
\bae{\label{eq: general fNL}
    \fNL(k_\uL,k_\uS)=\frac{5}{12}\frac{B_\zeta(k_\uL,k_\uS,k_\uS)}{P_\zeta(k_\uL)P_\zeta(k_\uS)},
}
as a useful variable for the squeezed configuration $k_\uL\ll k_\uS$.

\subsection{Maldacena's consistency relation as the cosmological soft theorem}\label{sec: Maldacena CR}

Before going into the details of the specific model, we here review the so-called Maldacena's consistency relation for the squeezed bispectrum~\cite{Maldacena:2002vr}.
Making use of the cubic action described in the next subsection, Maldacena showed that in the single-field slow-roll inflation, the squeezed bispectrum is related to the power spectrum by the relation,
\bae{\label{eq: Maldacena CR}
    B_\zeta(k_1,k_2,k_3)\underset{k_1/k_2\ll1}{\to}-\dv{\ln\calP_\zeta(k_2)}{\ln k_2}P_\zeta(k_1)P_\zeta(k_2).
}
It is now understood as a kind of cosmological soft theorem (see, e.g., Refs.~\cite{Creminelli:2004yq,Chen:2006dfn,Cheung:2007sv,Li:2008gg,Seery:2008ax,Urakawa:2009gb,Leblond:2010yq,Tanaka:2011aj,Creminelli:2011rh,Creminelli:2012ed,Hinterbichler:2012nm,Assassi:2012zq,Pajer:2013ana,Kenton:2016abp,Tada:2016pmk,Finelli:2017fml,Suyama:2020akr,Suyama:2021adn}).

Let us assume that on a large scale, the curvature perturbation $\zeta_\uL$ is time-independent and the metric is expressed as
\bae{
    \dd{s^2}=-\dd{t^2}+a^2(t)\ee^{2\zeta_\uL(\bfx)}\dd{\bfx^2}+\calO\pqty{\frac{k_\uL}{aH}},
}
where $k_\uL$ is the wavenumber associated with $\zeta_\uL$.
$\zeta_\uL$ is further renormalized into the rescaling of the spatial coordinate for a local universe:
\bae{
    \bar{\bfx}=\ee^{\zeta_\uL(\bfx=\mathbf{0})}\bfx,
}
which reduces the local metric to the background one,
\bae{
    \dd{s^2}=-\dd{t^2}+a^2(t)\dd{\bar{x}^2}+\calO(k_\uL^2\bar{x}^2)+\calO\pqty{\frac{k_\uL}{aH}}.
}
Therefore, if there is no intrinsic correlation between the long- and short-wavelength modes, the short modes are physically decoupled from $\zeta_\uL$.
Only apparent correlation arises from the coordinate transformation of the scalar curvature:
\bae{
    \hat{\zeta}_\uS(\bfx)=\hat{\bar{\zeta}}_\uS\bqty{\bar{\bfx}(\bfx)}\simeq\pqty{1+\zeta_\uL(0)x_i\partial_{x_i}}\hat{\bar{\zeta}}_\uS(\bfx),
}
or in Fourier space
\bae{
    \hat{\zeta}_{\bfk_\uS}\simeq\bqty{1-\zeta_\uL(0)(3+k_{\uS i}\partial_{k_{\uS i}})}\hat{\bar{\zeta}}_{\bfk_\uS}.
}
Hence the short-wavelength power spectrum (in the global comoving coordinate) is corrected by $\zeta_\uL$ as\footnote{Note that $k_\uS$ and $k_\uS'$ derivatives act also on the momentum conservation $(2\pi)^3\delta^{(3)}(\bfk_\uS+\bfk_\uS')$. Its derivative is understood in a Fourier-transform way as
\bae{
    (k_{\uS i}\partial_{k_{\uS i}}+k_{\uS i}'\partial_{k_{\uS i}'})(2\pi)^3\delta^{(3)}(\bfk_{\uS}+\bfk_\uS')&=(k_{\uS i}\partial_{k_{\uS i}}+k_{\uS i}'\partial_{k_{\uS i}'})\int\dd[3]{\bfx}\ee^{-i(\bfk_\uS+\bfk_\uS')\cdot\bfx} \nonumber \\
    &=\int\dd[3]{\bfx}\pqty{-i(\bfk_\uS+\bfk_\uS')\cdot\bfx}\ee^{-i(\bfk_\uS+\bfk_\uS')\cdot\bfx} \nonumber \\
    &=\int\dd[3]{\bfx}x_i\partial_{x_i}\ee^{-i(\bfk_\uS+\bfk_\uS')\cdot\bfx} \nonumber \\
    &=-3(2\pi)^3\delta^{(3)}(\bfk_\uS+\bfk_\uS'),
}
with use of the integration by parts in the last equation.}
\bae{
    \braket{\hat{\zeta}_{\bfk_\uS}\hat{\zeta}_{\bfk_\uS^\prime}}&\simeq\zeta_\uL(0)\bqty{-6-k_{\uS i}\partial_{k_{\uS i}}-k_{\uS i}^\prime\partial_{k_{\uS i}^\prime}}\braket{\hat{\bar{\zeta}}_{\bfk_\uS}\hat{\bar{\zeta}}_{\bfk_\uS^\prime}} \nonumber \\
    &=-(2\pi)^3\delta^{(3)}(\bfk_\uS+\bfk_\uS^\prime)\zeta_\uL(0)(3+k_{\uS i}\partial_{k_{\uS i}})P_{\bar{\zeta}}(k_\uS) \nonumber \\
    &=-(2\pi)^3\delta^{(3)}(\bfk_\uS+\bfk_\uS^\prime)\zeta_\uL(0)P_{\bar{\zeta}}(k_\uS)\dv{\ln\calP_{\bar{\zeta}}(k_\uS)}{\ln k_\uS} \nonumber \\
    &\simeq-(2\pi)^3\delta^{(3)}(\bfk_\uS+\bfk_\uS^\prime)\zeta_\uL(0)P_\zeta(k_\uS)\dv{\ln\calP_\zeta(k_\uS)}{\ln k_\uS}.
}
Therefore, the squeezed bispectrum has its correlation with $\zeta_\uL$ reproduces the consistency relation~\eqref{eq: Maldacena CR}.
Inversely speaking, the realization of Maldacena's consistency relation is the indicator of \emph{no physical correlation} between the long- and short-wavelength modes.
In this case, \acp{PBH} are expected to be not biased.
Making use of this consistency relation, Ref.~\cite{Pimentel:2012tw} also showed that the superhorizon curvature perturbations are conserved against the one-loop correction from the short-wavelength modes in the single-field slow-roll inflation.

\subsection{Cubic action and the Feynman rule}

The full action of the system is given by
\bae{
	S=\int\dd[4]{x}\sqrt{-g}\bqty{\frac{1}{2}R-\frac{1}{2}\partial_\mu\phi\partial^\mu\phi-V(\phi)},
}
where $R$ is the Ricci curvature of the spacetime metric, $\phi$ is the inflaton field, and $V(\phi)$ is its potential.
Taking the comoving gauge $\phi=\phi(t)$ (no perturbation in the inflaton field), making use of the \ac{ADM} formalism for the metric,
\bae{
	\dd{s^2}=-N^2\dd{t^2}+\gamma_{ij}(\dd{x^i}+\beta^i\dd{t})(\dd{x^j}+\beta^j\dd{t}),
}
where $N$ is the lapse function, $\beta^i$ is the shift vector, and $\gamma_{ij}$ is the 3-dim. metric, and neglecting the tensor mode as
\bae{
	\gamma_{ij}=a^2\ee^{2\zeta}\delta_{ij},
}
one obtains the action for the comoving curvature perturbation $\zeta$.

The terms cubic-order in $\zeta$ in the action are summarized as~\cite{Collins:2011mz,Arroja:2011yj,Adshead:2013zfa,Passaglia:2018afq,Passaglia:2018ixg}
\bae{
	S^{(3)}=S^{(3)}_\bulk+S^{(3)}_\EoM+S^{(3)}_\uB,
}
where
\beae{\label{eq: three S3}
	&S^{(3)}_\bulk=\bmte{\int\dd[4]{x}\left[a^3\epsilon^2\zeta\dot{\zeta}^2+a\epsilon^2\zeta(\partial\zeta)^2-2a\epsilon\dot{\zeta}(\partial\zeta)(\partial\chi)+\frac{a^3\epsilon}{2}\dot{\eta}\zeta^2\dot{\zeta} \right. \\
	\left.+\frac{\epsilon}{2a}(\partial\zeta)(\partial\chi)\partial^2\chi+\frac{\epsilon}{4a}(\partial^2\zeta)(\partial\chi)^2\right],} \\
	&S^{(3)}_\EoM=\int\dd[4]{x}2f(\zeta)\eval{\fdv{L}{\zeta}}_1, \\
	&S^{(3)}_\uB=\bmte{\int\dd[4]{x}\dv{t}\left[-9a^3H\zeta^3+\frac{a}{H}\zeta(\partial\zeta)^2-\frac{1}{4aH^3}(\partial\zeta)^2\partial^2\zeta-\frac{a\epsilon}{H}\zeta(\partial\zeta)^2-\frac{a^3\epsilon}{H}\zeta\dot{\zeta}^2 \right. \\
	\left. +\frac{1}{2aH^2}\zeta\qty(\partial_i\partial_j\zeta\partial_i\partial_j\chi-\partial^2\zeta\partial^2\chi)-\frac{a\eta}{2}\zeta^2\partial^2\chi-\frac{1}{2aH}\zeta\pqty{\partial_i\partial_j\chi\partial_i\partial_j\chi-\partial^2\chi\partial^2\chi}\right],}
}
with
\beae{
	&\chi=a^2\epsilon\partial^{-2}\dot{\zeta} \qc
	\eval{\fdv{L}{\zeta}}_1=a\pqty{\partial^2\dot{\chi}+H\partial^2\chi-\epsilon\partial^2\zeta}, \\
	&f(\zeta)=\bmte{\frac{\eta}{4}\zeta^2+\frac{1}{H}\zeta\dot{\zeta}+\frac{1}{4a^2H^2}\bqty{-(\partial\zeta)^2+\partial^{-2}\pqty{\partial_i\partial_j(\partial_i\zeta\partial_j\zeta)}} \\ 
	+\frac{1}{2a^2H}\bqty{(\partial\zeta)(\partial\chi)-\partial^{-2}\pqty{\partial_i\partial_j(\partial_i\zeta\partial_j\chi)}}.}
}
We dropped the spatial boundary terms to obtain this expression but kept the temporal boundary terms, which are actually relevant to the squeezed bispectrum.
$\partial$ represents the spatial derivative and $\partial^{-2}$ denotes the inverse Laplacian.
The linear \ac{EoM} for $\zeta$ is described as
\bae{
	\eval{\fdv{L}{\zeta}}_1=0.
}

One way to calculate the bispectrum from this cubic action is to utilize the field redefinition~\cite{Maldacena:2002vr,Arroja:2011yj},
\bae{\label{zetan}
	\zeta=\zeta_\un+f(\zeta_\un).
}
This redefinition in the quadratic action cancels $S^{(3)}_\EoM$ and $S^{(3)}_\uB$ and hence the cubic action for $\zeta_\un$ is much more simple as
\bae{
	S^{(3)}[\zeta_\un]=S^{(3)}_\bulk[\zeta_\un].
}
The bispectrum of $\zeta_\un$ is then related to that of $\zeta$ at the leading order in $\calP_\zeta$ by
\bae{\label{eq: zetan to zeta in bispectrum}
	\braket{\hat{\zeta}(\bfx_1)\hat{\zeta}(\bfx_2)\hat{\zeta}(\bfx_3)}\simeq\braket{\hat{\zeta}_\un(\bfx_1)\hat{\zeta}_\un(\bfx_2)\hat{\zeta}_\un(\bfx_3)}+\braket{\hat{\zeta}_\uI(\bfx_1)\hat{\zeta}_\uI(\bfx_2)f(\hat{\zeta}_\uI(\bfx_3))}+\text{(perms.)}.
}

Another way which we adopt in this section is to directly use the original cubic action~\eqref{eq: three S3}.
We hereafter consider the leading-order terms $\propto\epsilon$ in the $\epsilon$ expansion.
Among the bulk terms, only one term is relevant (note that $\eta$ is not necessarily small contrary to $\epsilon$ in the constant-roll models):
\bae{
	S^{(3)}_\bulk\simeq\int\dd{\tau}\dd[3]{\bfx}\frac{a^2\epsilon\eta^\prime}{2}\zeta^2\zeta^\prime,
}
where we changed the time variable to the conformal one $\tau$.
The \ac{EoM} term does not give any contribution because $\eval{\delta L/\delta\zeta}_1$ vanishes when the linear mode function is substituted in the calculation of the bispectrum.
As we will see below, the equal-time retarded propagator must be included in the contributions from the boundary terms. The equal-time retarded propagator for the same operator vanishes, so terms without $\zeta^\prime=\partial_\tau\zeta$ do not contribute to $\zeta$'s bispectrum.
Also, we are interested in the squeezed limit $k_1\ll k_2$ at the time when at least $k_1$ are well superhorizon.
In this case, the other momenta satisfy $\bfk_2\simeq-\bfk_3$ and then the sixth and the last terms in $S^{(3)}_\uB$ are suppressed in any momentum configuration.
Therefore, only the following two terms are relevant in the boundary action:
\bae{
	S^{(3)}_\uB\simeq\int\dd{\tau}\dd[3]{\bfx}\dv{\tau}\bqty{-\frac{a\epsilon}{H}\zeta{\zeta^\prime}^2-\frac{a^2\epsilon\eta}{2}\zeta^2\zeta^\prime}.
}

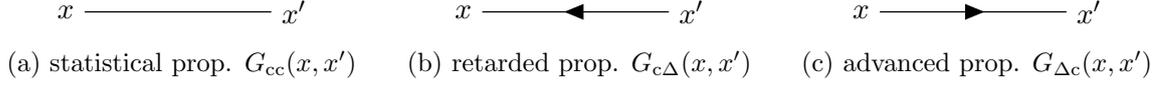
\begin{figure}
    \centering
    \begin{tabular}{c}
        \begin{minipage}{0.33\hsize}
            \centering
            \begin{tikzpicture}
                \begin{feynhand}
                    \vertex[particle] (a) at (-1.5,0) {$x$};
                    \vertex[particle] (b) at (1.5,0) {$x^\prime$};
                    \propag[plain] (a) to (b);
                \end{feynhand}
            \end{tikzpicture}
            \subcaption{statistical prop. $G_{\uc\uc}(x,x^\prime)$}
        \end{minipage}
        \begin{minipage}{0.33\hsize}
            \centering
            \begin{tikzpicture}
                \begin{feynhand}
                    \vertex[particle] (a) at (-1.5,0) {$x$};
                    \vertex[particle] (b) at (1.5,0) {$x^\prime$};
                    \propag[fermion] (b) to (a);
                \end{feynhand}
            \end{tikzpicture}
            \subcaption{retarded prop. $G_{\uc\Delta}(x,x^\prime)$}
        \end{minipage}
        \begin{minipage}{0.33\hsize}
            \centering
            \begin{tikzpicture}
                \begin{feynhand}
                    \vertex[particle] (a) at (-1.5,0) {$x$};
                    \vertex[particle] (b) at (1.5,0) {$x^\prime$};
                    \propag[fermion] (a) to (b);
                \end{feynhand}
            \end{tikzpicture}
            \subcaption{advanced prop. $G_{\Delta\uc}(x,x^\prime)$}
        \end{minipage}
    \end{tabular}
    \caption{Diagrams for propagators.}
    \label{fig: propagator}
\end{figure}

One can formulate Feynman's diagrammatic rule for these interactions in the Schwinger--Keldysh picture (see, e.g., Ref.~\cite{2009AdPhy..58..197K} for a review and Refs.~\cite{Calzetta:1986ey,Tsamis:1996qq,Tsamis:1996qm,Weinberg:2005vy,vanderMeulen:2007ah,Seery:2007we,Seery:2008ax,Prokopec:2010be,Leblond:2010yq,Gong:2016qpq,Chen:2016nrs,Chen:2017ryl,Tokuda:2017fdh,Tokuda:2018eqs} for its application to cosmology).
There, the path integral is defined along the time path $C$ from the sufficient past $\tau_{-\infty}$ to the sufficient future $\tau_\infty$ and then again back to $\tau_{-\infty}$.
The curvature perturbation is doubled as $\zeta_+$ and $\zeta_-$ living in the forward and backward paths respectively and they are connected by the boundary condition $\zeta_+(\tau_\infty)=\zeta_-(\tau_\infty)$.
The tree-level propagator is defined by the time-ordered two-point function along $C$ as
\bae{
	G_{ab}(x,x^\prime)&=\braket{T_C\hat{\zeta}_{a\uI}(x)\hat{\zeta}_{b\uI}(x^\prime)} \nonumber \\
	&=\bce{
		\Theta(\tau-\tau^\prime)\braket{\hat{\zeta}_\uI(x)\hat{\zeta}_\uI(x^\prime)}+\Theta(\tau^\prime-\tau)\braket{\hat{\zeta}_\uI(x^\prime)\hat{\zeta}_\uI(x)}, & (a,b)=(+,+), \\
		\braket{\hat{\zeta}_\uI(x^\prime)\hat{\zeta}_\uI(x)}, & (a,b)=(+,-), \\
		\braket{\hat{\zeta}_\uI(x)\hat{\zeta}_\uI(x^\prime)}, & (a,b)=(-,+), \\
		\Theta(\tau^\prime-\tau)\braket{\hat{\zeta}_\uI(x)\hat{\zeta}_\uI(x^\prime)}+\Theta(\tau-\tau^\prime)\braket{\hat{\zeta}_\uI(x^\prime)\hat{\zeta}_\uI(x)}, & (a,b)=(-,-),
	}
}
where $\Theta$ is the step function
\bae{
	\Theta(x)=\bce{
		1, & x>0, \\
		1/2, & x=0, \\
		0, & x<0.
	}
}
One often redefines the field basis as
\bae{
	\zeta_\uc=\frac{\zeta_++\zeta_-}{2} \qc \zeta_\Delta=\zeta_+-\zeta_-,
}
called the Schwinger--Keldysh basis. On this basis, the propagator is rewritten as
\bae{
	G_{\alpha\beta}(x,x^\prime)=\bce{
		\dps
		\frac{1}{2}\braket{\{\hat{\zeta}_\uI(x),\hat{\zeta}_\uI(x^\prime)\}}, & (\alpha,\beta)=(\uc,\uc), \\
		\Theta(\tau-\tau^\prime)\braket{[\hat{\zeta}_\uI(x),\hat{\zeta}_\uI(x^\prime)]}, & (\alpha,\beta)=(\uc,\Delta), \\
		\Theta(\tau^\prime-\tau)\braket{[\hat{\zeta}_\uI(x^\prime),\hat{\zeta}_\uI(x)]}, & (\alpha,\beta)=(\Delta,\uc), \\
		0, & (\alpha,\beta)=(\Delta,\Delta),
	}
}
which are illustrated by lines with or without an arrow as shown in Fig.~\ref{fig: propagator}.
$G_{\uc\uc}$, $G_{\uc\Delta}$, and $G_{\Delta\uc}$ are referred to as statistical, retarded, and advanced Green's functions, respectively.
Obviously, the identities
\bae{
    G_{\uc\uc}(x,x^\prime)=G_{\uc\uc}(x^\prime,x) \qc
    G_{\uc\Delta}(x,x^\prime)=G_{\Delta\uc}(x^\prime,x),
}
follow their definitions.
One can also include the momentum $\zeta^\prime=\partial_\tau\zeta$ in the propagator as
\bege{\label{eq: momentum propagator}
    G_{\alpha\bar{\beta}}(x,x^\prime)=\braket{T_C\hat{\zeta}_{\alpha\uI}(x)\hat{\zeta}^\prime_{\beta\uI}(x^\prime)} \qc
    G_{\bar{\alpha}\beta}(x,x^\prime)=\braket{T_C\hat{\zeta}_{\alpha\uI}^\prime(x)\hat{\zeta}_{\beta\uI}(x^\prime)}, \\
    G_{\bar{\alpha}\bar{\beta}}=\braket{T_C\hat{\zeta}_{\alpha\uI}^\prime(x)\hat{\zeta}_{\beta\uI}^\prime(x^\prime)},
}
illustrated as Fig.~\ref{fig: momentum propagator}. 
In Fourier space, one finds
\bae{
	G_{\alpha\beta}(\tau,\tau^\prime;k)=\bce{
		\Re\zeta_k(\tau)\zeta_k^*(\tau^\prime), & (\alpha,\beta)=(\uc,\uc), \\
		2i\Theta(\tau-\tau^\prime)\Im\zeta_k(\tau)\zeta_k^*(\tau^\prime), & (\alpha,\beta)=(\uc,\Delta), \\
		-2i\Theta(\tau^\prime-\tau)\Im\zeta_k(\tau)\zeta_k^*(\tau^\prime), & (\alpha,\beta)=(\Delta,\uc), \\
		0, & (\alpha,\beta)=(\Delta,\Delta).
	}
}
Propagators including momentums are similarly defined.
We note that the equal-time statistical propagator for $\zeta$ is equivalent to the ordinary power spectrum, $G_{\uc\uc}(\tau,\tau;k)=P_\zeta(\tau,k)$, and the equal-time retarded one for $\zeta$ and $\zeta^\prime$ is determined by the Wronskian condition as $G_{\uc\bar{\Delta}}(\tau,\tau;k)=i/(4a^2\epsilon)$ (note that $\Theta(0)=1/2$) independently of $k$. The equal-time retarded propagator for the same operator vanishes.

\begin{figure}
    \centering
    \begin{tabular}{c}
        \begin{minipage}{0.33\hsize}
            \centering
            \begin{tikzpicture}
                \begin{feynhand}
                    \vertex[particle] (a) at (-1.5,0) {$x$};
                    \vertex (b) at (0,0);
                    \vertex[particle] (c) at (1.5,0) {$x^\prime$};
                    \propag[plain] (a) to (b);
                    \propag[sca] (b) to (c);
                \end{feynhand}
            \end{tikzpicture}
            \subcaption{$G_{\uc\bar{\uc}}(x,x^\prime)$}
        \end{minipage}
        \begin{minipage}{0.33\hsize}
            \centering
            \begin{tikzpicture}
                \begin{feynhand}
                    \vertex[particle] (a) at (-1.5,0) {$x$};
                    \vertex[particle] (b) at (1.5,0) {$x^\prime$};
                    \propag[sca] (a) to (b);
                \end{feynhand}
            \end{tikzpicture}
            \subcaption{$G_{\bar{\uc}\bar{\uc}}(x,x^\prime)$}
        \end{minipage} \\
        \begin{minipage}{0.33\hsize}
            \centering
            \begin{tikzpicture}
                \begin{feynhand}
                    \vertex[particle] (a) at (-1.5,0) {$x$};
                    \vertex (b) at (0.5,0);
                    \vertex[particle] (c) at (1.5,0) {$x^\prime$};
                    \propag[fermion] (b) to (a);
                    \propag[sca] (b) to (c);
                \end{feynhand}
            \end{tikzpicture}
            \subcaption{$G_{\uc\bar{\Delta}}(x,x^\prime)$}
        \end{minipage}
        \begin{minipage}{0.33\hsize}
            \centering
            \begin{tikzpicture}
                \begin{feynhand}
                    \vertex[particle] (a) at (-1.5,0) {$x$};
                    \vertex (b) at (-0.5,0);
                    \vertex[particle] (c) at (1.5,0) {$x^\prime$};
                    \propag[sca] (a) to (b);
                    \propag[fermion] (c) to (b);
                \end{feynhand}
            \end{tikzpicture}
            \subcaption{$G_{\bar{\uc}\Delta}(x,x^\prime)$}
        \end{minipage}
        \begin{minipage}{0.33\hsize}
            \centering
            \begin{tikzpicture}
                \begin{feynhand}
                    \vertex[particle] (a) at (-1.5,0) {$x$};
                    \vertex[particle] (b) at (1.5,0) {$x^\prime$};
                    \propag[chasca] (b) to (a);
                \end{feynhand}
            \end{tikzpicture}
            \subcaption{$G_{\bar{\uc}\bar{\Delta}}(x,x^\prime)$}
        \end{minipage}
    \end{tabular}
    \caption{Propagators including the momentum $\zeta^\prime$~\eqref{eq: momentum propagator}.}
    \label{fig: momentum propagator}
\end{figure}

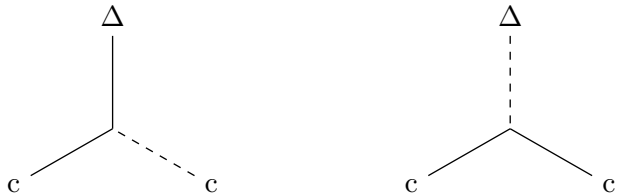
\begin{figure}
	\centering
	\begin{tabular}{c}
		\begin{minipage}{0.33\hsize}
			\centering
			\begin{tikzpicture}
				\begin{feynhand}
					\vertex[particle] (a) at (0,1.5) {$\Delta$};
					\vertex[particle] (b) at (-1.3,-0.75) {$\uc$};
					\vertex[particle] (c) at (1.3,-0.75) {$\uc$};
					\vertex (d) at (0,0);
					\propag[plain] (a) to (d);
					\propag[plain] (b) to (d);
					\propag[sca] (c) to (d);
				\end{feynhand}
			\end{tikzpicture}
		\end{minipage}
		\begin{minipage}{0.33\hsize}
			\centering
			\begin{tikzpicture}
				\begin{feynhand}
					\vertex[particle] (a) at (0,1.5) {$\Delta$};
					\vertex[particle] (b) at (-1.3,-0.75) {$\uc$};
					\vertex[particle] (c) at (1.3,-0.75) {$\uc$};
					\vertex (d) at (0,0);
					\propag[sca] (a) to (d);
					\propag[plain] (b) to (d);
					\propag[plain] (c) to (d);
				\end{feynhand}
			\end{tikzpicture}
		\end{minipage}
	\end{tabular}
	\caption{The two main couplings in the bulk cubic Lagrangian. The vertex value $ia^2\epsilon\eta^\prime$ is assigned to both vertices.}
	\label{fig: bulk coupling}
\end{figure}

Recalling the action in the Schwinger--Keldysh formalism is given by $S=S[\zeta_+]-S[\zeta_-]$ where the minus sign of $S[\zeta_-]$ comes from the backward time flow for $\zeta_-$, one finds the cubic bulk Lagrangian as
\bae{
	\calL^{(3)}_\bulk&=\frac{a^2\epsilon\eta^\prime}{2}(\zeta_+^2\zeta_+^\prime-\zeta_-^2\zeta_-^\prime) \nonumber \\
	&=\frac{a^2\epsilon\eta^\prime}{2}\pqty{2\zeta_\uc\zeta_\Delta\zeta_\uc^\prime+\zeta_\uc^2\zeta_\Delta^\prime+\frac{1}{4}\zeta_\Delta^2\zeta_\Delta^\prime},
}
where the first two terms practically give dominant contributions because the retarded (or advanced) propagator necessarily takes the non-dominant mode in the mode function contrary to the statistical propagator. The interchange of two $\zeta_\uc$'s for the second term always yields the factor $2$, so in the Feynman rule, the vertex value $ia^2\epsilon\eta^\prime$ is assigned both to the first and second terms as summarized in Fig.~\ref{fig: bulk coupling}.
In our case, $\eta^\prime$ yields the Dirac deltas as
\bae{
    \eta^\prime=\Delta\eta(\tau_\us)\delta(\tau-\tau_\us)+\Delta\eta(\tau_\ue)\delta(\tau-\tau_\ue) \qc \Delta\eta(\tau_\us)=-\Delta\eta(\tau_\ue)=2\beta.
}
The time integration is hence simplified.

\begin{figure}
	\centering
	\begin{tabular}{c}
	\begin{minipage}{0.25\hsize}
		\centering
		\begin{tikzpicture}
			\begin{feynhand}
				\vertex[particle] (a) at (0,1.5) {$\Delta$};
				\vertex[particle] (b) at (-1.3,-0.75) {$\uc$};
				\vertex[particle] (c) at (1.3,-0.75) {$\uc$};
				\vertex[ringdot] (d) at (0,0) {};
				\propag[plain] (a) to (d);
				\propag[plain] (b) to (d);
				\propag[sca] (c) to (d);
			\end{feynhand}
		\end{tikzpicture}
	\end{minipage}
	\begin{minipage}{0.25\hsize}
		\centering
		\begin{tikzpicture}
			\begin{feynhand}
				\vertex[particle] (a) at (0,1.5) {$\Delta$};
				\vertex[particle] (b) at (-1.3,-0.75) {$\uc$};
				\vertex[particle] (c) at (1.3,-0.75) {$\uc$};
				\vertex[ringdot] (d) at (0,0) {};
				\propag[sca] (a) to (d);
				\propag[plain] (b) to (d);
				\propag[plain] (c) to (d);
			\end{feynhand}
		\end{tikzpicture}
	\end{minipage} 
	\begin{minipage}{0.25\hsize}
		\centering
		\begin{tikzpicture}
			\begin{feynhand}
				\vertex[particle] (a) at (0,1.5) {$\Delta$};
				\vertex[particle] (b) at (-1.3,-0.75) {$\uc$};
				\vertex[particle] (c) at (1.3,-0.75) {$\uc$};
				\vertex[dot] (d) at (0,0) {};
				\propag[plain] (a) to (d);
				\propag[sca] (b) to (d);
				\propag[sca] (c) to (d);
			\end{feynhand}
		\end{tikzpicture}
	\end{minipage}
	\begin{minipage}{0.25\hsize}
		\centering
		\begin{tikzpicture}
			\begin{feynhand}
				\vertex[particle] (a) at (0,1.5) {$\Delta$};
				\vertex[particle] (b) at (-1.3,-0.75) {$\uc$};
				\vertex[particle] (c) at (1.3,-0.75) {$\uc$};
				\vertex[dot] (d) at (0,0) {};
				\propag[sca] (a) to (d);
				\propag[sca] (b) to (d);
				\propag[plain] (c) to (d);
			\end{feynhand}
		\end{tikzpicture}
	\end{minipage}
	\end{tabular}
	\caption{Cubic vertices valued $-2i\delta(\tilde{\tau}-\tau)a^2\epsilon\eta$ (white dot) and $-4i\delta(\tilde{\tau}-\tau)\frac{a\epsilon}{H}$ (black dot) coming from the boundary action $S^{(3)}_\uB$. $\tilde{\tau}$ is the time of the pair of the $\Delta$ mode.}
	\label{fig: B vertex}
\end{figure}
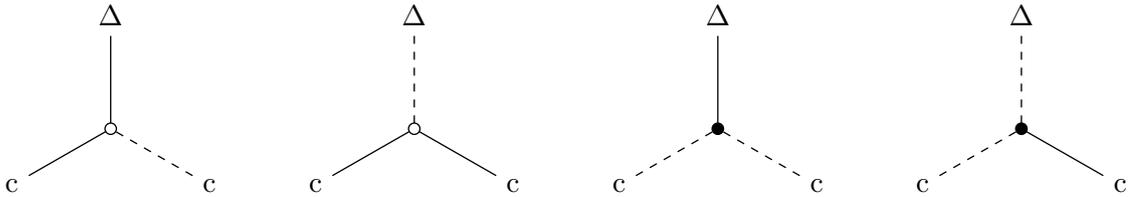

The Feynman rule for the boundary terms is understood as follows. Expressing the (dominant) boundary Lagrangian as $\calL^{(3)}_\uB=\dv{\tau}\calB$ where each term in $\calB$ includes one $\Delta$ mode,
one would calculate expectation values such as $\braket{T_C\hat{\calO}_\uI(\tau_1,\tau_2,\tau_3,\cdots)\dv{\tau}\hat{\calB}_\uI(\tau)}$ and integrate it over $\tau$. Here $\hat{\calO}_\uI$ and $\hat{\calB}_\uI$ are products in the interaction picture and $\hat{\calO}_\uI$ can be characterized by several times $\tau_1$, $\tau_2$, $\tau_3$, $\cdots$.
Let us then clarify the difference between $\braket{T_C\hat{\calO}_\uI(\tau_1,\tau_2,\tau_3,\cdots)\dv{\tau}\hat{\calB}_\uI(\tau)}$ and $\dv{\tau}\braket{T_C\hat{\calO}_\uI(\tau_1,\tau_2,\tau_3,\cdots)\hat{\calB}_\uI(\tau)}$.
It comes from the time ordering, i.e., the time derivative of the step function from the retarded or advanced propagator.
Hence one finds the relation (note that $2i\Im\zeta_k(\tau)\zeta_k^*(\tau)=2G_{\uc\Delta}(\tau,\tau;k)$ due to $\Theta(0)=1/2$),
\bme{
	\Braket{T_C\hat{\calO}(\tau_1,\tau_2,\tau_3,\cdots)\dv{\tau}\hat{\calB}_\uI(\tau)} \nonumber \\
	=\dv{\tau}\braket{T_C\hat{\calO}(\tau_1,\tau_2,\tau_3,\cdots)\hat{\calB}_\uI(\tau)}+2\delta(\tau_1-\tau)G_{\uc\Delta}(\tau_1,\tau)G_{\uc\uc}(\tau_2,\tau)G_{\uc\uc}(\tau_3,\tau)\cdots + \text{(perms.)},
}
where propagators in the last line can include the momentum $\zeta^\prime$. The first term on the right-hand side gives the boundary contributions at $\tau_{-\infty}$ and $\tau_\infty$. 
The $\tau_{-\infty}$ boundary is dropped by the $-i\epsilon$ prescription similarly to the ordinary interacting theory~\cite{Maldacena:2002vr}, while the $\tau_\infty$ is prohibited due to the retarded propagator.
Therefore, the vertex is practically understood as $\propto\delta(\tilde{\tau}-\tau)$ where $\tilde{\tau}$ is the other time of the arguments of the retarded or advanced propagator.
The concrete values are summarized in Fig.~\ref{fig: B vertex}.
By connecting them and imposing the momentum conservation at each vertex, one can calculate the squeezed bispectrum as we concretely see in the following subsections.

\subsection{Bispectrum after the constant-roll phase}

\begin{figure}
	\centering
	\begin{tabular}{c}
		\begin{minipage}{0.33\hsize}
			\centering
			\begin{tikzpicture}
				\begin{feynhand}
					\vertex (a) at (-1.5,1);
					\vertex (b) at (0.5,1);
					\vertex (c) at (1.5,1);
					\vertex (d) at (0,-1);
                    \vertex (e) at (0.25,0);
					\propag[plain] (d) to [out=135, in=-90, edge label=$k_\uL$] (a);
                    \propag[sca] (d) to (e);
                    \propag[plain] (e) to [edge label=$k_\uS$] (b);
					\propag[fermion] (d) to [out=45, in=-90, edge label'=$k_\uS^\prime$] (c);
				\end{feynhand}
			\end{tikzpicture}
		\end{minipage}
		\begin{minipage}{0.33\hsize}
			\centering
			\begin{tikzpicture}
				\begin{feynhand}
					\vertex (a) at (-1.5,1);
					\vertex (b) at (0.5,1);
					\vertex (c) at (1.5,1);
					\vertex (d) at (0,-1);
                    \vertex (e) at (0.25,0);
					\propag[plain] (d) to [out=135, in=-90, edge label=$k_\uL$] (a);
                    \propag[sca] (d) to (e);
					\propag[fermion] (e) to [edge label=$k_\uS$] (b);
					\propag[plain] (d) to [out=45, in=-90, edge label'=$k_\uS^\prime$] (c);
				\end{feynhand}
			\end{tikzpicture}
		\end{minipage}
	\end{tabular}
	\caption{Two leading diagrams for the squeezed bispectrum evaluated in the deep slow-roll phase.}
	\label{fig: bispectrum in SR}
\end{figure}
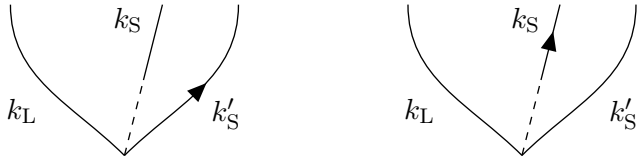

\begin{figure}
	\centering
	\begin{tabular}{c}
		\begin{minipage}[b]{0.5\hsize}
			\centering
			\includegraphics[width=0.95\hsize]{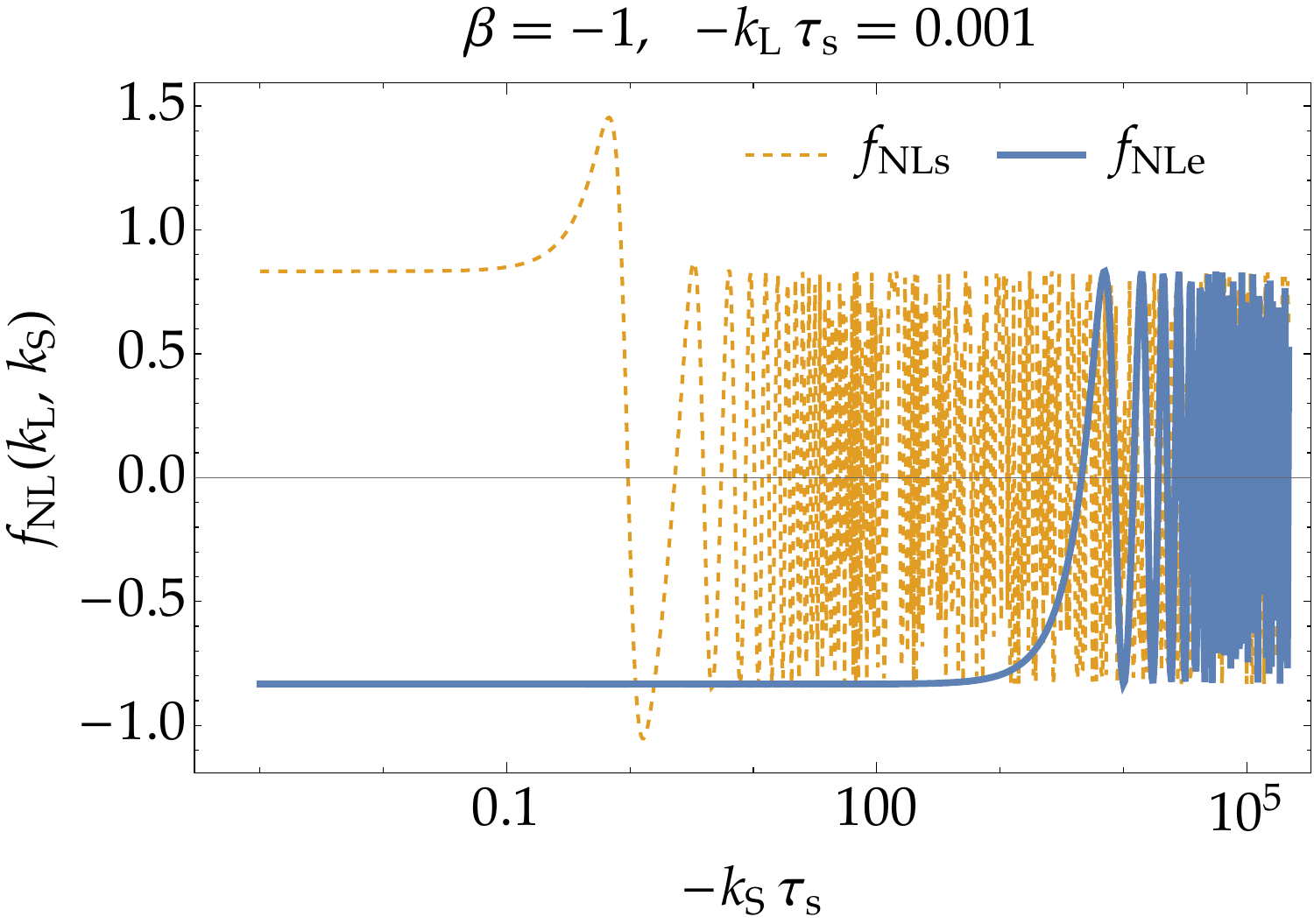}
		\end{minipage}
		\begin{minipage}[b]{0.5\hsize}
			\centering
			\includegraphics[width=0.95\hsize]{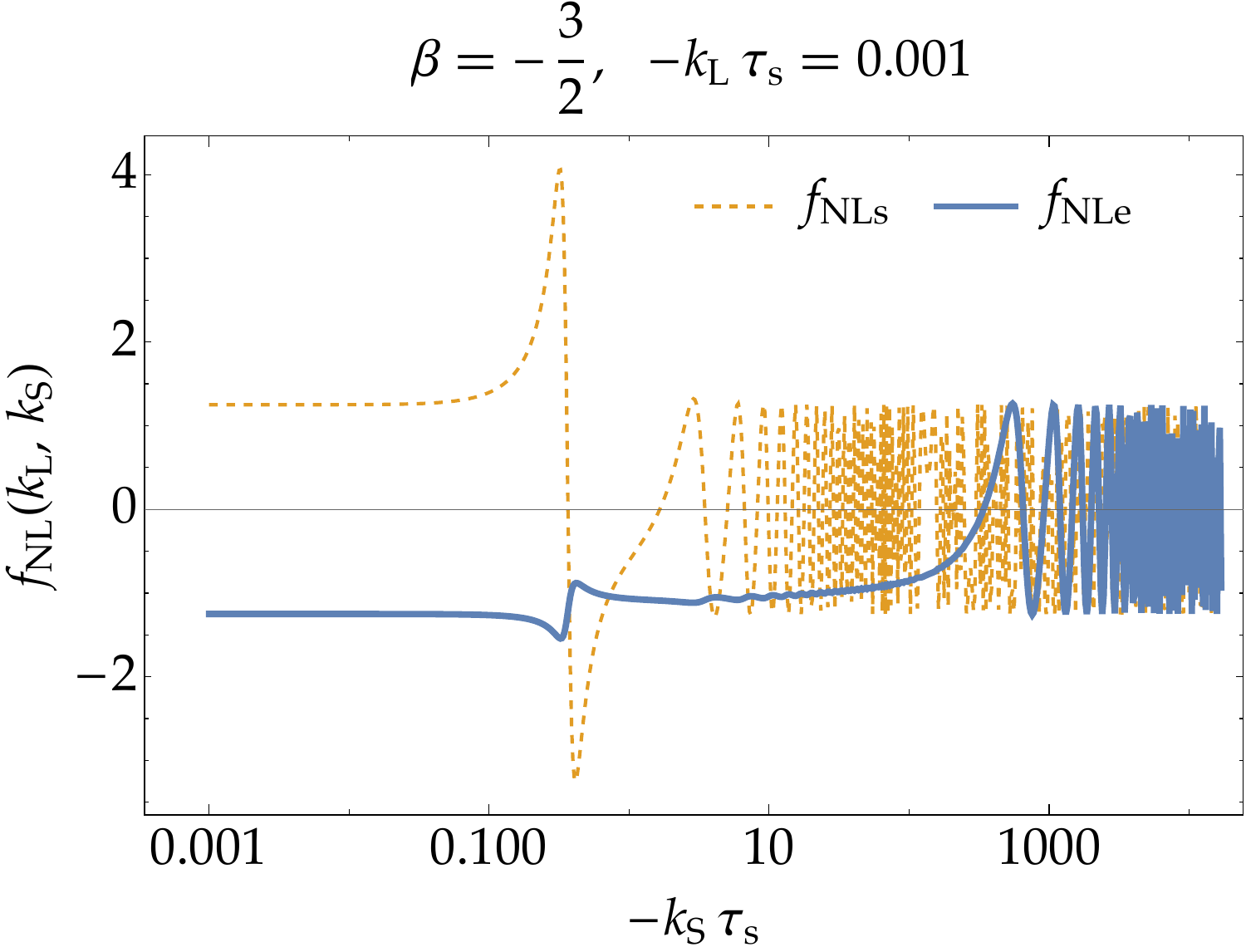}
		\end{minipage} \\
		\begin{minipage}{0.5\hsize}
			\centering
			\includegraphics[width=0.95\hsize]{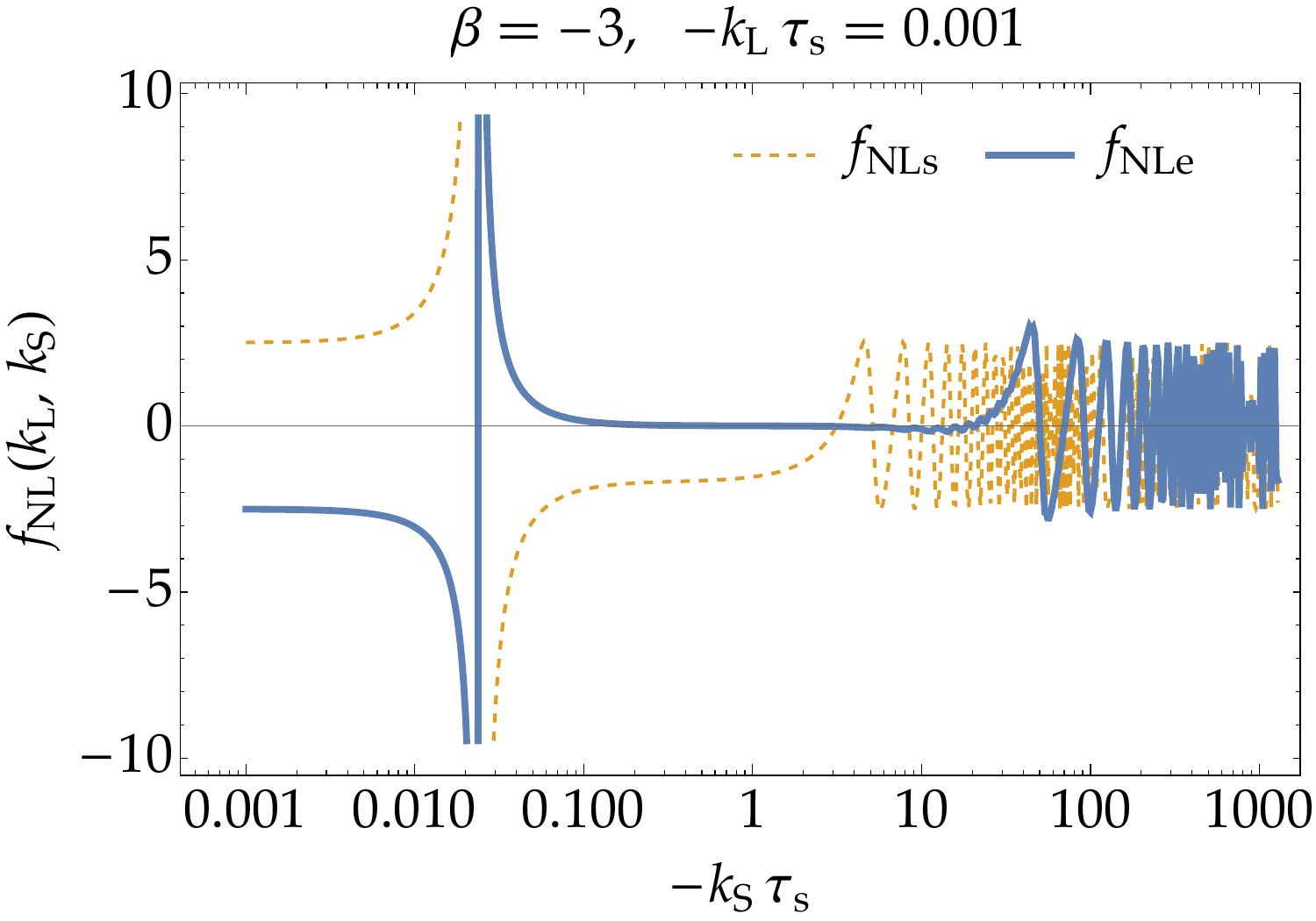}
		\end{minipage}
	\end{tabular}
	\caption{The bispectrum contributions from $\tau_\us$ (orange dashed) and $\tau_\ue$ (blue) in terms of the generalized non-linearity parameter $\fNL(k_\uL,k_\uS)$~\eqref{eq: general fNL} evaluated at sufficiently late time $\tau\to0$ for $\beta=-1$ (top-left), $-3/2$ (top-right), and $-3$ (bottom) with $k_\uL=-0.001/\tau_\us$.}
	\label{fig: fNLs and fNLe}
\end{figure}

Let us first investigate the squeezed bispectrum evaluated in the deep second slow-roll phase well after the constant-roll phase.
There, both $\eta$ and $\zeta^\prime$ have been decayed away and hence the boundary terms summarized in Fig.~\ref{fig: B vertex} can be neglected.
One helpful rule is that diagrams including the statistical propagator of the long mode $k_\uL$ dominate in the squeezed limit because the dimensionful power spectrum is inversely proportional to the Fourier volume factor: $P_\zeta(k_\uL)\propto k_\uL^{-3}$.
The main contributions are hence given by the two diagrams shown in Fig.~\ref{fig: bispectrum in SR}.
Each term is doubled due to the interchange $k_\uS\leftrightarrow k_\uS^\prime$ and has the contributions from $\tau_\us$ and $\tau_\ue$.
Therefore, one finds
\bae{
    B_\zeta(k_\uL,k_\uS,k_\uS)&\bmte{=2ia^2(\tau_\us)\epsilon(\tau_\us)\Delta\eta(\tau_\us)G_{\uc\uc}(\tau,\tau_\us;k_\uL)\left[G_{\uc\bar{\uc}}(\tau,\tau_\us;k_\uS)G_{\uc\Delta}(\tau,\tau_\us;k_\uS) \right. \\
    \left.+G_{\uc\bar{\Delta}}(\tau,\tau_\us;k_\uS)G_{\uc\uc}(\tau,\tau_\us;k_\uS)\right]+\text{($\tau_\us\to\tau_\ue$)}} \nonumber \\
    &=-4a^2(\tau_\us)\epsilon(\tau_\us)\Delta\eta(\tau_\us)\Re\bqty{\zeta_{k_\uL}(\tau)\zeta_{k_\uL}^*(\tau_\us)}\Im\bqty{\zeta_{k_\uS}^2(\tau)\zeta_{k_\uS}^*(\tau_\us)\zeta_{k_\uS}^{*\prime}(\tau_\us)}+\text{($\tau_\us\to\tau_\ue$)},
}
where the second term is obtained by replacing $\tau_\us$ in the first term by $\tau_\ue$.
In Fig.~\ref{fig: fNLs and fNLe}, we show an example bispectrum in the form of the generalized non-linearity parameter~\eqref{eq: general fNL}. One sees that the contributions from $\tau_\us$ and $\tau_\ue$ are the same order of magnitude.

\begin{figure}
    \centering
    \begin{tabular}{c}
        \begin{minipage}[b]{0.5\hsize}
            \centering
            \includegraphics[width=0.95\hsize]{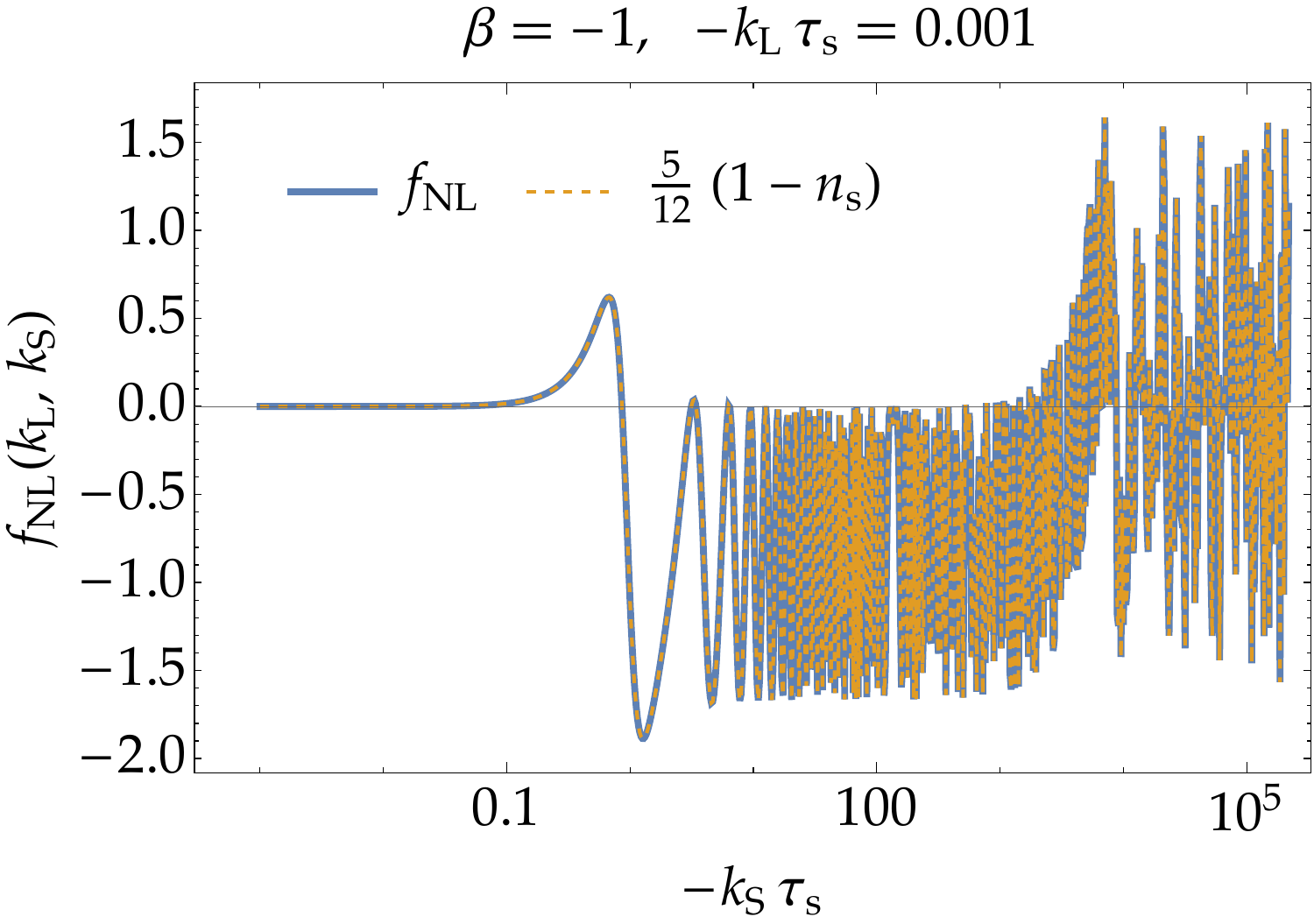}
        \end{minipage}
        \begin{minipage}[b]{0.5\hsize}
            \centering
            \includegraphics[width=0.95\hsize]{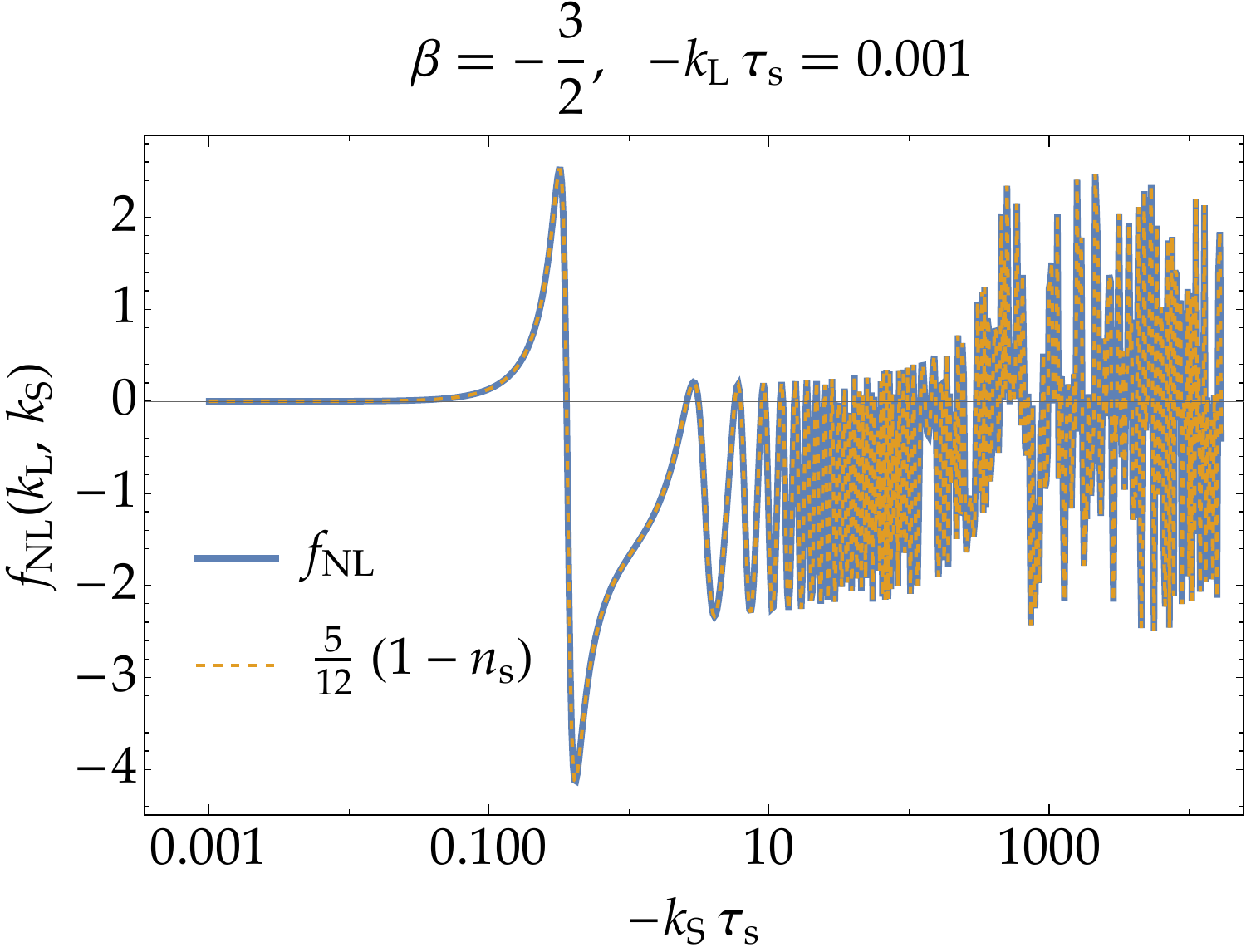}
        \end{minipage} \\ 
        \begin{minipage}{0.5\hsize}
            \centering
            \includegraphics[width=0.95\hsize]{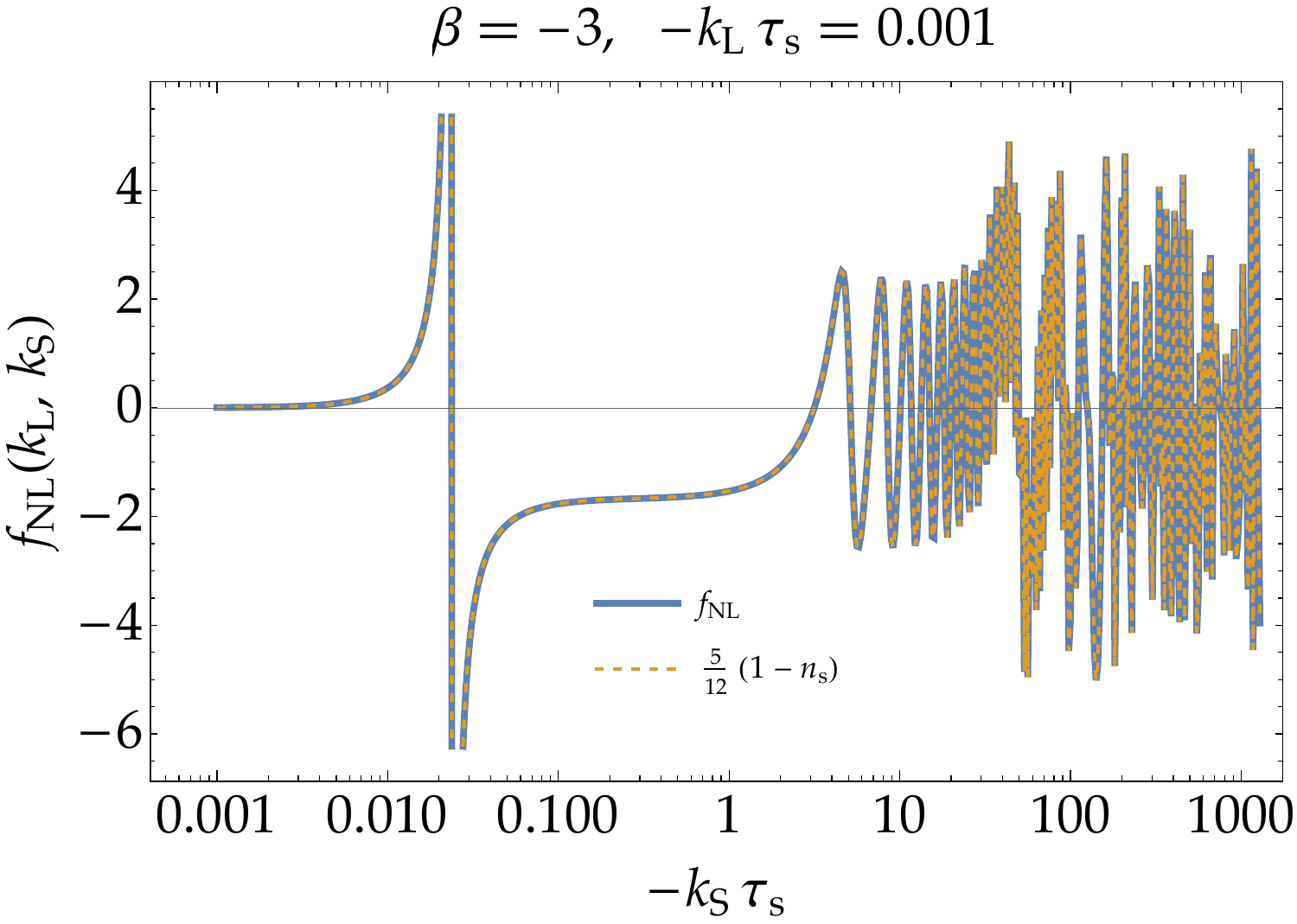}
        \end{minipage}
    \end{tabular}
    \caption{The total non-linearity parameter $\fNL(k_\uL,k_\uS)$ with $k_\uL=-0.001/\tau_\us$ (blue) and the spectral index $\frac{5}{12}\qty(1-\ns(k_\uS))$ (orange dashed) evaluated at sufficiently late time $\tau\to0$ for $\beta=-1$ (top-left), $-3/2$ (top-right), and $-3$ (bottom). For every model, Maldacena's consistency relation~\eqref{eq: CR in fNL} holds well.}
    \label{fig: fNL ks SR}
\end{figure}

In Fig.~\ref{fig: fNL ks SR}, we compare the non-linearity parameter with the spectral index for the short-wavelength mode: 
\bae{
    \ns(k_\uS)-1=\dv{\ln\calP_\zeta(k_\uS)}{\ln k_\uS}.
}
One finds that for a sufficiently long-wavelength mode $-k_\uL\tau_\us\ll1$ which exits the horizon well before the onset of the constant-roll phase, Maldacena's consistency relation
\bae{\label{eq: CR in fNL}
    \fNL(k_\uL,k_\uS)\underset{\text{CR}}{=}\frac{5}{12}\qty(1-\ns(k_\uS)),
}
holds well for any value of $\beta$ and for any short-wavelength mode $k_\uS$ either on superhorizon or subhorizon scales at the end of the constant-roll phase.
This result is consistent with the intuition: as $\zeta_{k_\uL}$ is well frozen after exiting the horizon (even during or after the constant-roll phase), the logic developed in Sec.~\ref{sec: Maldacena CR} can be applied and $\zeta_{k_\uL}$ is merely viewed as a (comoving) scale shift by the short-wavelength modes.
Physics in each local patch cannot distinguish $\zeta_{k_\uL}$. \acp{PBH} produced in this model will hence show no spatial modulation beyond the adiabatic perturbation on a large scale~\cite{Suyama:2020akr}.

\begin{figure}
    \centering
    \begin{tabular}{c}
        \begin{minipage}[b]{0.5\hsize}
            \centering
            \includegraphics[width=0.95\hsize]{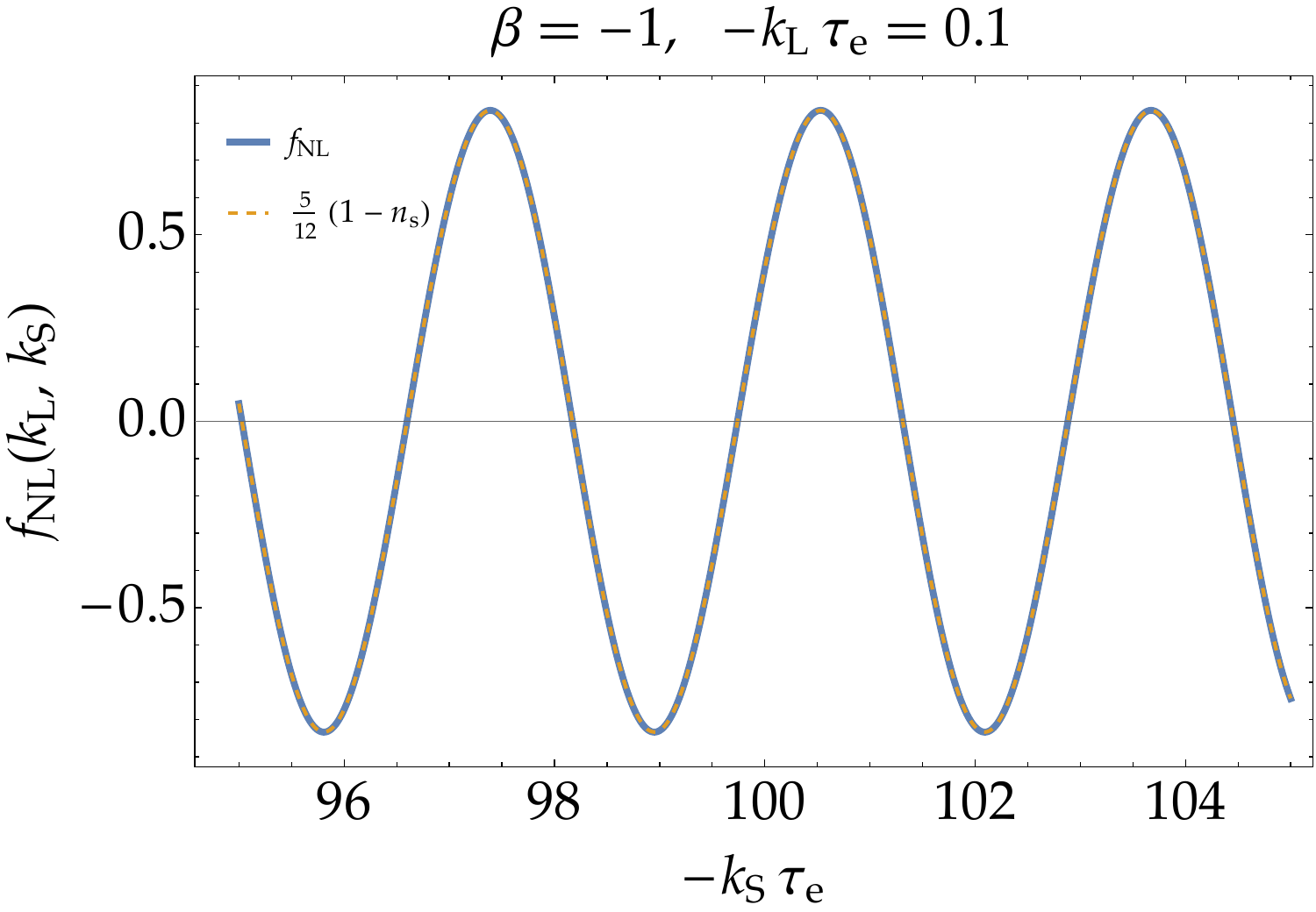}
        \end{minipage}
        \begin{minipage}[b]{0.5\hsize}
            \centering
            \includegraphics[width=0.95\hsize]{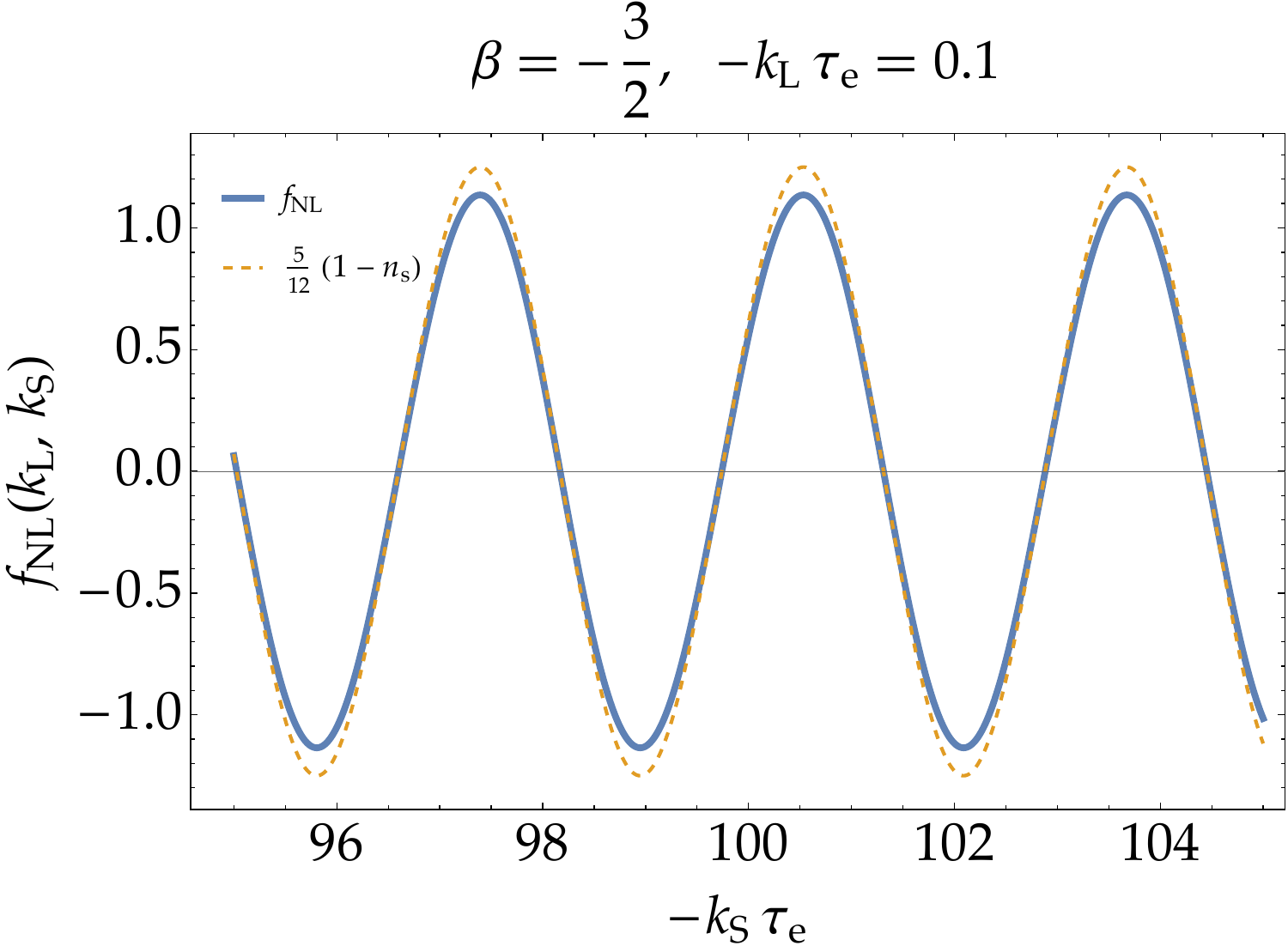}
        \end{minipage} \\
        \begin{minipage}{0.5\hsize}
            \centering
            \includegraphics[width=0.95\hsize]{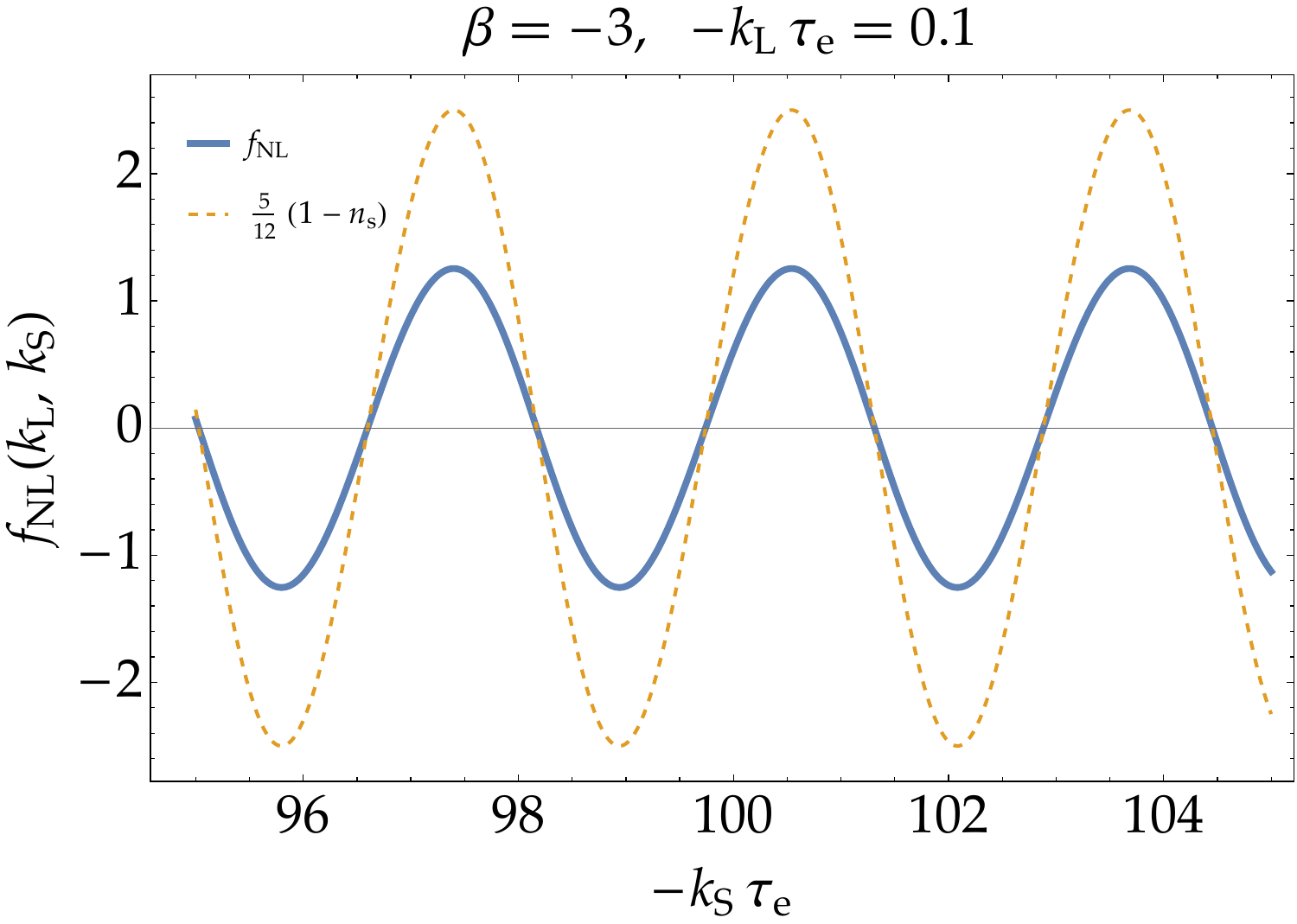}
        \end{minipage}
        \begin{minipage}{0.5\hsize}
        	\centering
			\includegraphics[width=0.95\hsize]{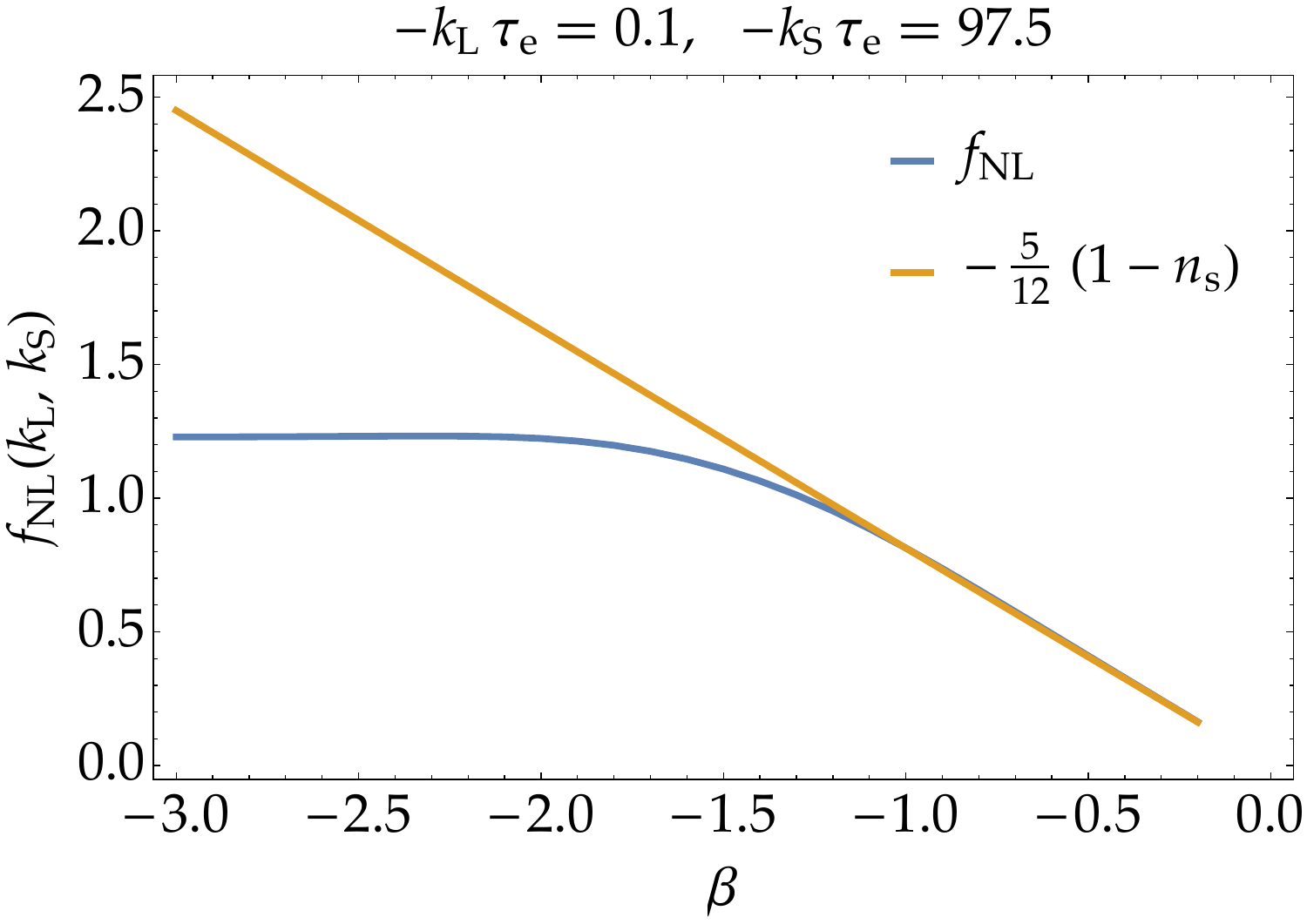}
        \end{minipage}
    \end{tabular}
    \caption{The similar plots to Fig.~\ref{fig: fNL ks SR} but for $k_\uL=-0.1/\tau_\ue$ (top and bottom-left). 
    The bottom-right panel compares $\fNL$ and $-5(1-\ns)/12$ for $k_\uS=-97.5/\tau_\ue$, which shows the violation of Maldacena's consistency relation for the non-attractor model $\beta\leq-3/2$.}
    \label{fig: fNL ke SR}
\end{figure}

On the other hand, for modes around the peak scale $\sim-1/\tau_\ue$, violations of the consistency relation can be found for the non-attractor model $\beta\leq-3/2$ as shown in Fig.~\ref{fig: fNL ke SR}.
This is because the curvature perturbation grows even on a superhorizon scale for the non-attractor model and it cannot be simply renormalized into the local scale factor.
The peak-scale curvature perturbation itself is hence non-Gaussian and the \ac{PBH} abundance is expected to be affected.
The practical computation of these bispectra is done by the mode function with the Bunch--Davies initial condition set during the constant-roll phase, neglecting the $\tau_\us$ contribution, because all relevant modes are well subhorizon at $\tau_\us$.

\subsection{Bispectrum during the constant-roll phase}

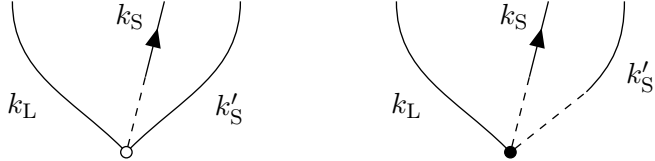
\begin{figure}
    \centering
    \begin{tabular}{c}
		\begin{minipage}{0.33\hsize}
			\centering
			\begin{tikzpicture}
				\begin{feynhand}
					\vertex (a) at (-1.5,1);
					\vertex (b) at (0.5,1);
					\vertex (c) at (1.5,1);
					\vertex[ringdot] (d) at (0,-1) {};
                    \vertex (e) at (0.25,0);
					\propag[plain] (d) to [out=135, in=-90, edge label=$k_\uL$] (a);
                    \propag[sca] (d) to (e);
					\propag[fermion] (e) to [edge label=$k_\uS$] (b);
					\propag[plain] (d) to [out=45, in=-90, edge label'=$k_\uS^\prime$] (c);
				\end{feynhand}
			\end{tikzpicture}
		\end{minipage}
        \begin{minipage}{0.33\hsize}
            \centering
            \begin{tikzpicture}
                \begin{feynhand}
                    \vertex (a) at (-1.5,1);
					\vertex (b) at (0.5,1);
					\vertex (c) at (1.5,1);
					\vertex[dot] (d) at (0,-1) {};
                    \vertex (e) at (0.25,0);
                    \vertex (f) at (1,-0.2);
					\propag[plain] (d) to [out=135, in=-90, edge label=$k_\uL$] (a);
                    \propag[sca] (d) to (e);
                    \propag[fermion] (e) to [edge label=$k_\uS$] (b);
                    \propag[sca] (d) to (f);
					\propag[plain] (f) to [out=45, in=-90, edge label'=$k_\uS^\prime$] (c);
                \end{feynhand}
            \end{tikzpicture}
        \end{minipage}
    \end{tabular}
    \caption{Two additional contributions from the boundary terms.}
    \label{fig: boundary for bispectrum}
\end{figure}

\begin{figure}
	\centering
	\begin{tabular}{c}
		\begin{minipage}[b]{0.5\hsize}
			\centering
			\includegraphics[width=0.95\hsize]{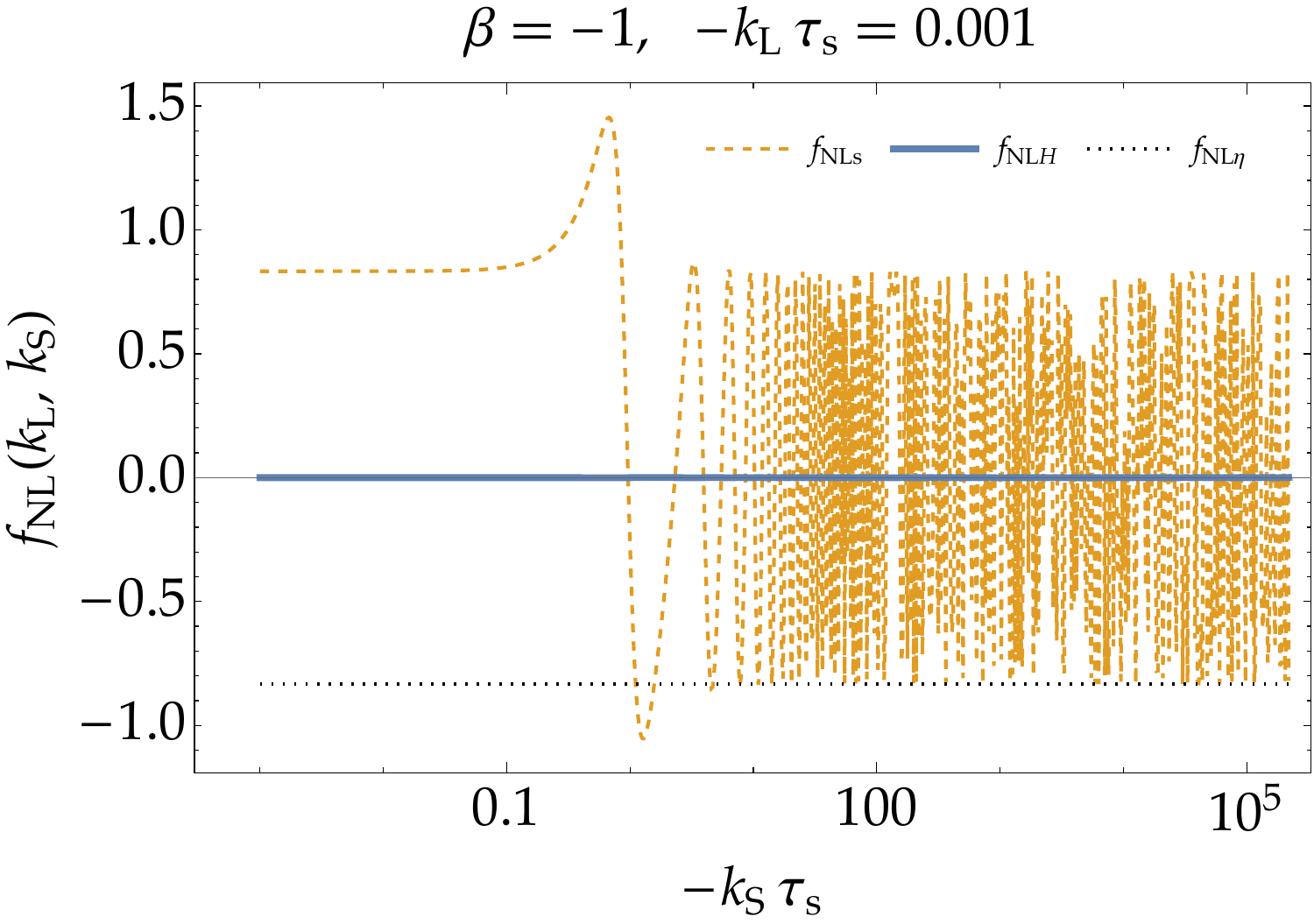}
		\end{minipage}
		\begin{minipage}[b]{0.5\hsize}
			\centering
			\includegraphics[width=0.95\hsize]{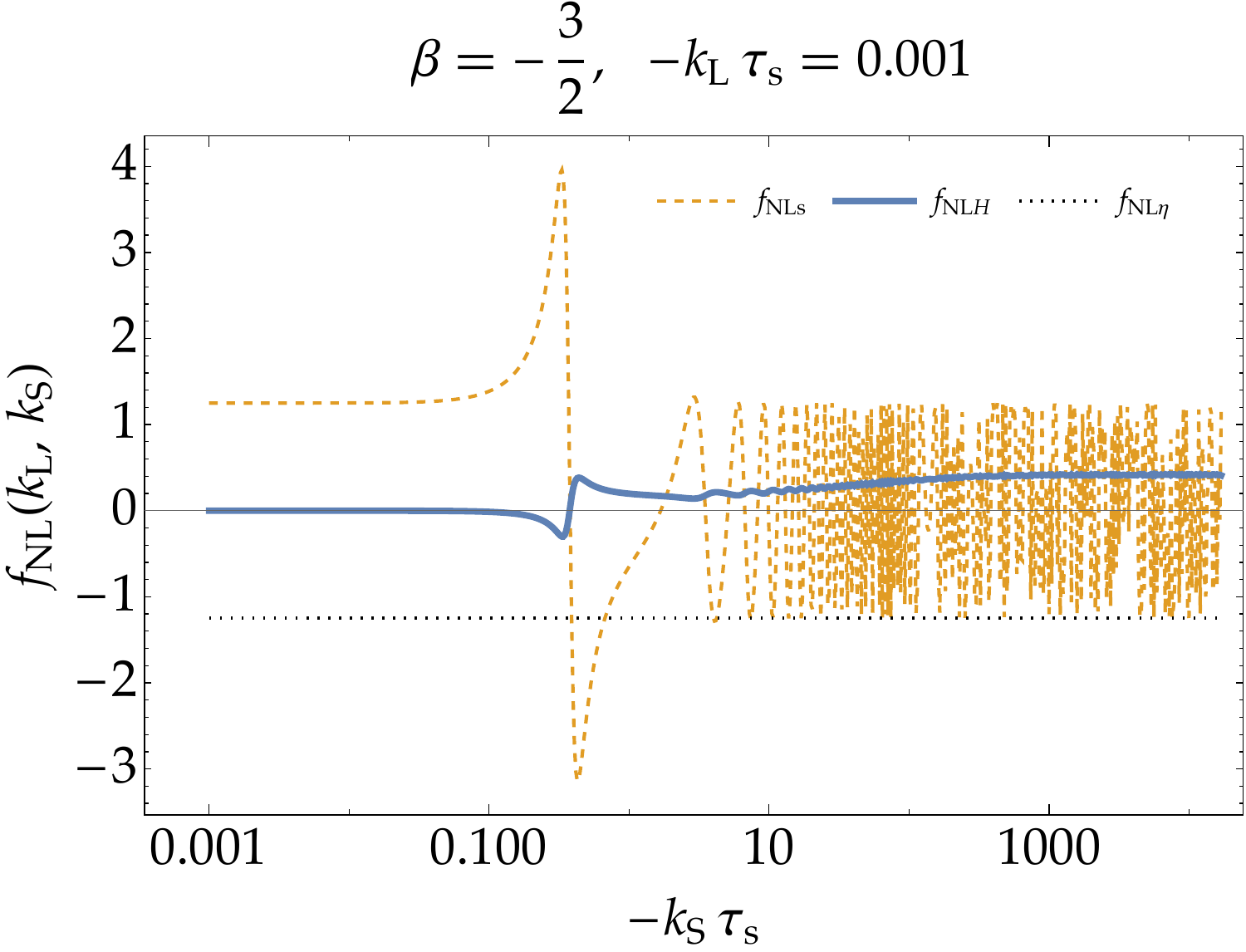}
		\end{minipage} \\
		\begin{minipage}{0.5\hsize}
			\centering
			\includegraphics[width=0.95\hsize]{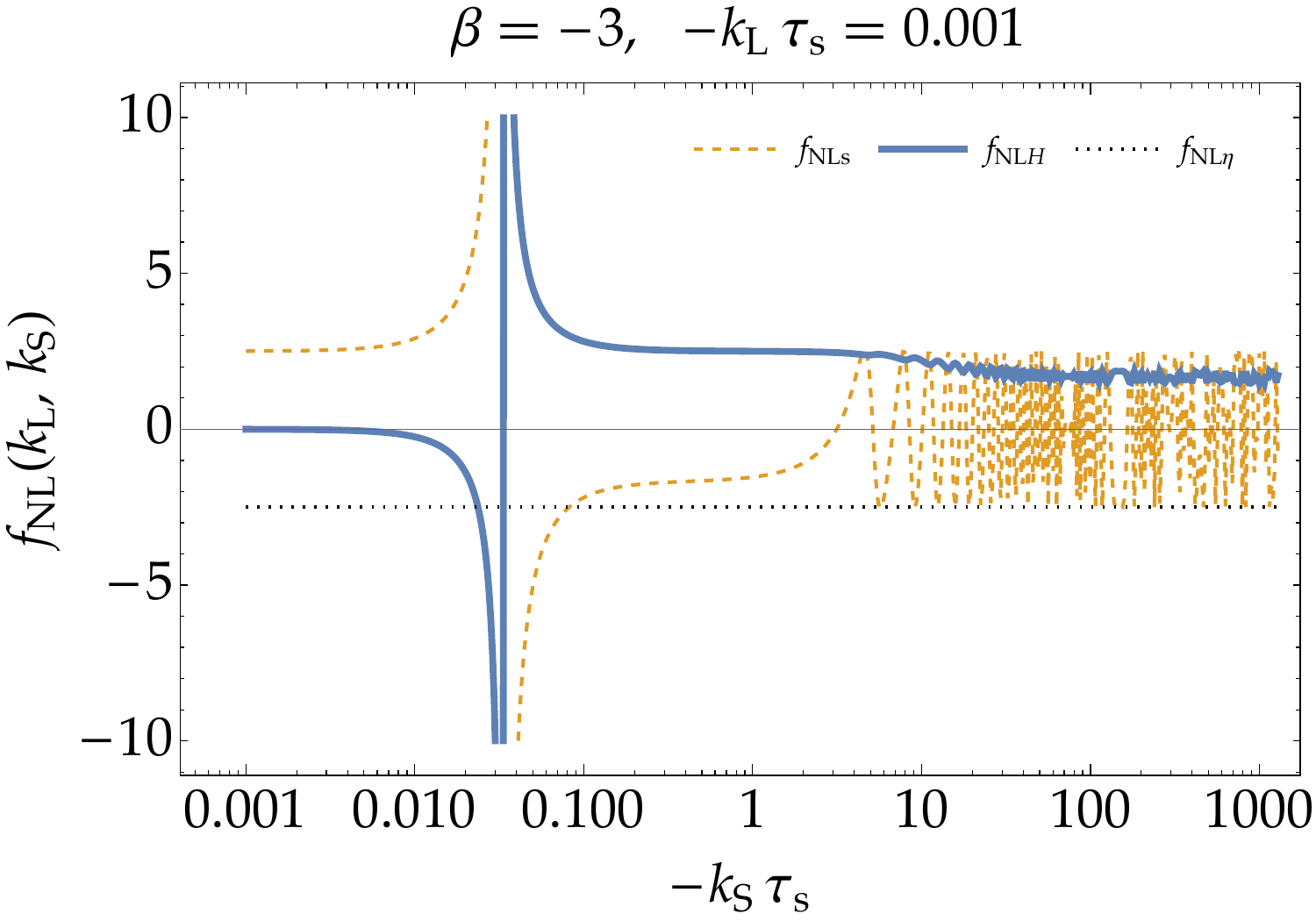}
		\end{minipage}
	\end{tabular}
	\caption{The bispectrum contributions from the bulk vertex at $\tau_\us$ (orange dashed), the $H$-boundary (blue), and $\eta$-boundary vertices (black dotted) evaluated during the constant-roll phase ($\tau\to\tau_\ue-0$) for $\beta=-1$ (top-left), $-3/2$ (top-right), and $-3$ (bottom) with $k_\uL=-0.001/\tau_\us$.}
	\label{fig: fNLs fNLeta fNLH}
\end{figure}

\begin{figure}
	\centering
	\begin{tabular}{c}
		\begin{minipage}[b]{0.5\hsize}
			\centering
			\includegraphics[width=0.95\hsize]{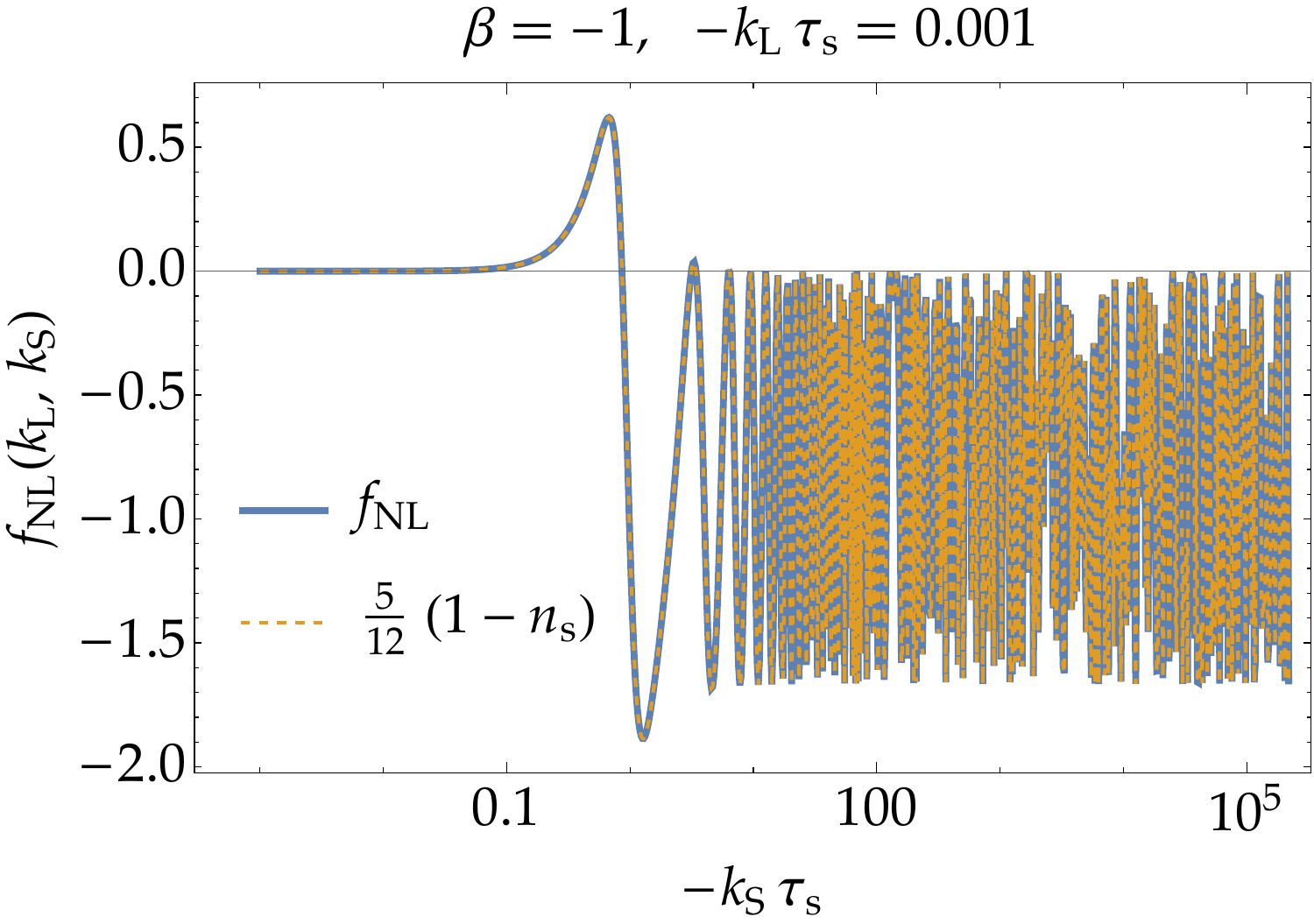}
		\end{minipage}
		\begin{minipage}[b]{0.5\hsize}
			\centering
			\includegraphics[width=0.95\hsize]{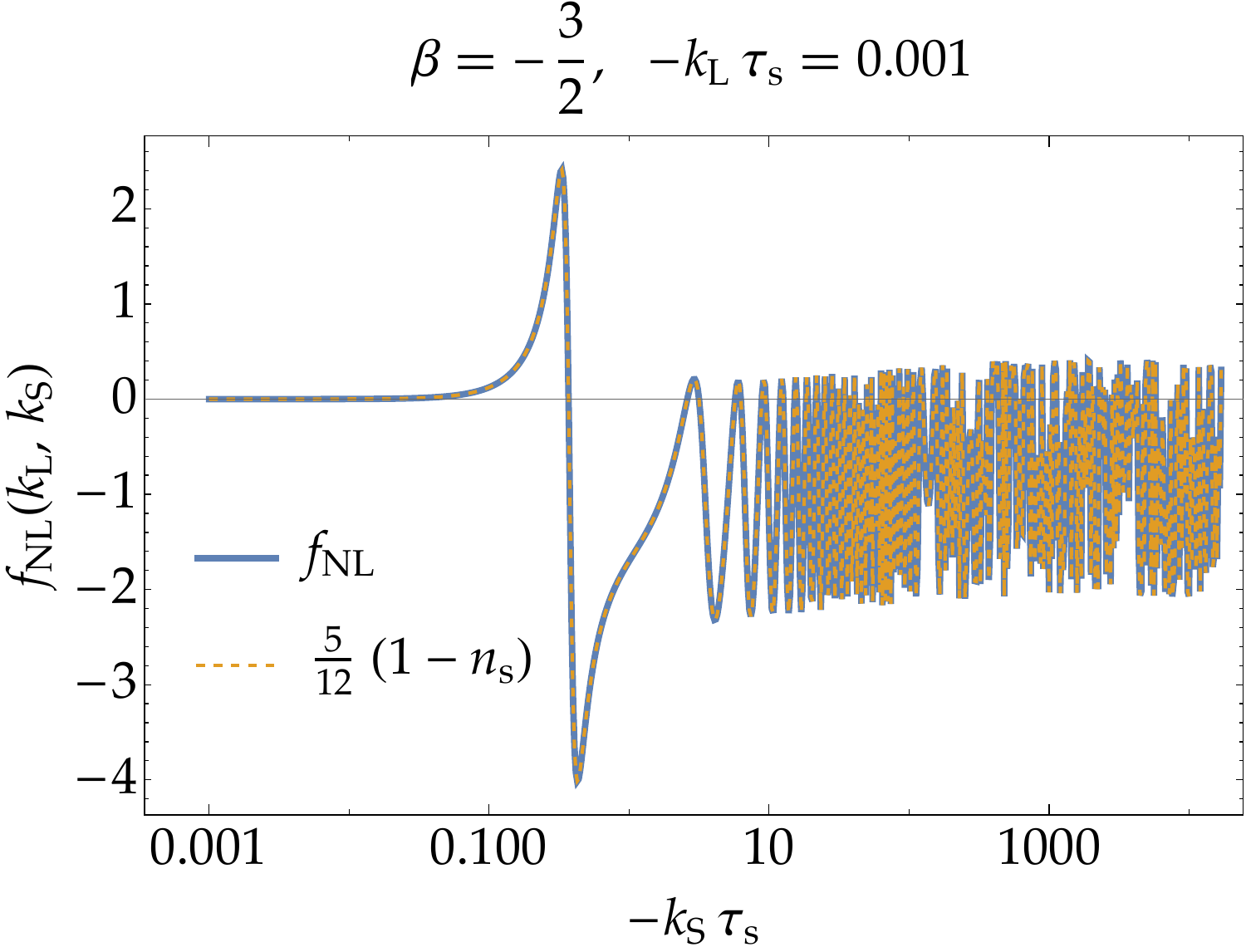}
		\end{minipage} \\
		\begin{minipage}{0.5\hsize}
			\centering
			\includegraphics[width=0.95\hsize]{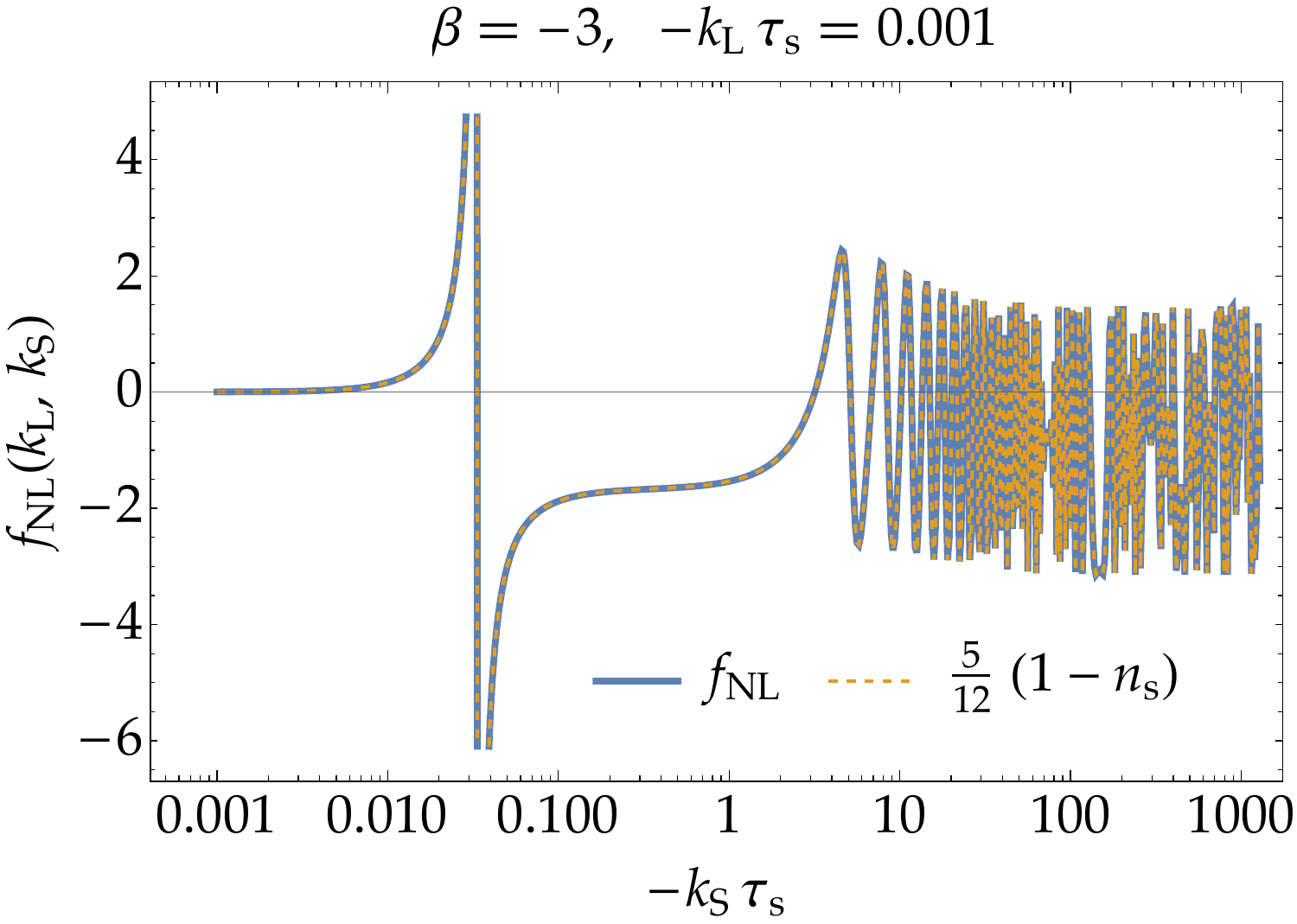}
		\end{minipage}
	\end{tabular}
	\caption{The total non-linearity parameter $\fNL(k_\uL,k_\uS)$ with $k_\uL=-0.001/\tau_\us$ (blue) and the spectral index $\frac{5}{12}\qty(1-\ns(k_\uS))$ (orange dashed) evaluated during the constant-roll phase ($\tau\to\tau_\ue-0$) for $\beta=-1$ (top-left), $-3/2$ (top-right), and $-3$ (bottom). Maldacena's consistency relation~\eqref{eq: CR in fNL} again holds well for every model thanks to the boundary terms.}
	\label{fig: fNLks CR}
\end{figure}

Let us also discuss the bispectrum evaluated during the constant-roll phase though it does not directly affect the \ac{PBH} physics.
Boundary terms cannot be neglected in this case.
In addition to the bulk contribution Fig.~\ref{fig: bispectrum in SR}, two diagrams shown in Fig.~\ref{fig: boundary for bispectrum} contribute, which we call the $\eta$-boundary and $H$-boundary terms, respectively (recall that the equal-time retarded propagator for the same operators vanishes for the $\eta$-boundary term).
In terms of the non-linearity parameter, they read
\beae{
    &\fNLeta=-\frac{5}{12}\times4ia^2\epsilon\eta G_{\uc\bar{\Delta}}(\tau,\tau;k_\uS)=\frac{5}{12}\eta, \\
    &\fNLH=-\frac{5}{12}\times8i\frac{a\epsilon}{H}\frac{G_{\uc\bar{\Delta}}(\tau,\tau;k_\uS)G_{\uc\bar{\uc}}(\tau,\tau;k_\uS)}{P_\zeta(k_\uS)}=\frac{5}{6}\frac{\Re[\zeta_{k_\uS}\zeta_{k_\uS}^{*\prime}]}{aHP_\zeta(k_\uS)},
}
where we used the Wronskian condition $G_{\uc\bar{\Delta}}(\tau,\tau;k)=i/(4a^2\epsilon)$. These contributions are equivalent to the redefinition contributions in Eq.~\eqref{eq: zetan to zeta in bispectrum} when using the $\zeta_\un$ approach as pointed out in Ref.~\cite{Arroja:2011yj}.
In Fig.~\ref{fig: fNLs fNLeta fNLH}, we exemplify these contributions. They are indeed non-negligible compared to the bulk vertex contribution $\fNLs$.
Fig.~\ref{fig: fNLks CR} shows that the total bispectrum satisfies Maldacena's consistency relation for a sufficiently long mode $k_\uL$ even during the constant-roll phase as expected because such a long mode is frozen enough even at the onset of the constant-roll phase.
Therefore, the boundary terms in the cubic action in terms of $\zeta$ are necessary for the realization of Maldacena's consistency relation.

\section{One-loop correction to the long-wavelength mode}
\label{sec:loop}

After elaborating on the squeezed bispectrum, now we are interested in the one-loop corrections in the transient constant-roll inflation.
To evaluate the one-loop contributions, it is necessary to take into account all relevant terms carefully, which is beyond the scope of the present paper. 
In this section, as a preliminary step, we focus on one-loop terms evaluated in Ref.~\cite{Kristiano:2022maq} for the ultra slow-roll scenario and clarify how the estimation is modified for a more general constant-roll scenario.
We shall see that the \ac{PBH} production in the transient constant-roll inflation is not prohibited from the perturbativity requirement on those one-loop terms.

Following Ref.~\cite{Kristiano:2022maq}, we calculate the one-loop 
correction to the power spectrum at $\tau_\ue$ and the \ac{CMB} pivot scale $k_*=\SI{0.05}{Mpc^{-1}}$ originating from 
the $\dot \eta$ term in $S^{(3)}_\bulk$ in Eq.~\eqref{eq: three S3}
at the instantaneous transition $\tau=\tau_\ue$
by using the field redefinition $\zeta\to\zeta_\un$ given in Eq.~\eqref{zetan} in the ordinary in-in formalism.
Before the computation, let us analytically solve the junction condition~\eqref{eq: junction}, making use of the asymptotic expansion of the Hankel functions~\cite{NIST:DLMF}
\begin{align}
	\label{subh} &\lim_{z\to\infty} H^{(1/2)}_\nu(z) \approx \sqrt{\f{2}{\pi z}} \ee^{\pm i \mk{z-\f{\pi}{4}(2\nu+1)}} \mk{ 1 \pm \f{i b}{z}} , \\
	\label{suph} &\lim_{z\to 0} H^{(1/2)}_\nu(z) \approx \mp \f{i}{\pi}\Gamma(\nu) \mk{\f{z}{2}}^{-\nu} , 
\end{align}
where 
\bae{
	b=\f{4\nu^2-1}{8}.
}
The curvature perturbation on subhorizon scales during the constant-roll phase follows Eq.~\eqref{subh} as
\bae{
	\zeta_k^\UV = \f{iH}{2 \Mpl \sqrt{\epsilon_{\SR1}}} \f{1}{k^{3/2}} F(k,\tau) ,
}
where 
\bae{ 
	F(k,\tau) = \mk{\f{\tau_\us}{\tau}}^\beta \kk{ - \Ck (b + i k \tau) \ee^{-i (k \tau + \pi (2 \nu + 1)/4)} 
	+ \Dk (b - i k \tau) \ee^{i (k \tau + \pi (2 \nu + 1)/4)}  }.
}
Here and henceforth, $\nu$ and $b$ denote $\nucr$ and $b_{\rm CR}$ for simplicity.
The connection at $\tau_\us$ deep inside the horizon is solved as
\beae{
	&\Ck= \f{i \beta b - (\beta + 1) (b - 1) k \tau_\us + i (\beta - 2 b + 2) k^2 \tau_\us^2 - 2 k^3 \tau_\us^3}{2 k \tau_\us [b (b - 1) + k^2 \tau_\us^2]} \ee^{i \pi (2 \nu + 1)/4} ,\\
	&\Dk= \f{-i \beta b + (\beta b + \beta - b + 1) k \tau_\us + i \beta k^2 \tau_\us^2}{2 k \tau_\us [b (b - 1) + k^2 \tau_\us^2]} \ee^{- i (\pi (2 \nu - 3)/4 + 2 k \tau_\us)} .
}
Using Eq.~\eqref{suph}, the power spectrum at $\tau_\ue$, $\calP_\CR(k)$, is amplified from the first slow-roll phase solution $\calP_\SR$ as 
\bae{ 
	\label{PCR} \calP_\CR(k)=\calP_\SR(k_\us)\abs{F(k,\tau_\ue)}^2 = \mk{\f{k_\us}{k_*}}^{-0.03} \calP_\SR(k_*) |F(k,\tau_\ue)|^2 .
}
Here, we assumed that $\calP_\SR$ is almost constant for $k\in[k_\us=-\tau_\us^{-1},k_\ue=-\tau_\ue^{-1}]$ but introduced a factor difference $\mk{k_\us/k_*}^{-0.03}$ between $k_\us$ and $k_*$.

The one-loop correction $\calP_{(1)}$ evaluated in Ref.~\cite{Kristiano:2022maq} is generalized to
\bae{
	\calP_{(1)}(k_*) = \f{1}{4} (\Delta\eta)^2 \calP_\SR^2(k_*) \mk{\f{k_\us}{k_*}}^{-0.03} \int_{k_\us}^{k_\ue} \f{\dd{k}}{k} |F(k,\tau_\ue)|^2,
}
where $\Delta\eta=-2\beta$ at $\tau=\tau_\ue$, and the factor $\mk{k_\us/k_*}^{-0.03}$ was introduced in \eqref{PCR}.
Here, the integral 
\bae{
	\label{integral-def} I \equiv \int_{k_\us}^{k_\ue} \f{\dd{k}}{k} |F(k,\tau_\ue)|^2 = \int_{1}^{\ell_\ue} \f{\dd{\ell}}{\ell} |F(\ell k_\us, \tau_\ue)|^2,
}
can be evaluated numerically, and the result is a function of $\ell_\ue\equiv k_\ue/k_\us$ for given $\beta$. 
An analytic form under the assumption $k_\ue/k_\us\gg 1$ is given by 
\beq \label{integral-app} I \simeq \mk{\f{k_\ue}{k_\us}}^{-2\beta} \mk{ c_0+c_1 \ln \f{k_\ue}{k_\us} } , \eeq
where 
\beae{
	&c_0= \bmte{\f{1}{2} + \f{1}{120} \beta (\beta+1)^2 (\beta+2)^2 
	\left[ \f{5}{16} ( 5 \beta^5 + 24 \beta^4 + 44 \beta^3 + 72 \beta^2 + 143 \beta + 72) \right. \\
	-\f{1}{8} (12 \beta^5 + 58 \beta^4 + 65 \beta^3 + 48 \beta^2 + 101 \beta + 36 ) \cos 2 \\
	\left. + \f{1}{2} (7 \beta^4 + 24 \beta^3 + 27 \beta^2 + 26 \beta - 4) \sin 2 \right], } \\
	&c_1= \f{1}{4} (\beta+1)^2 (\beta+2)^2 .
}
We provide a derivation of the expression~\eqref{integral-app} and a comparison with numerical calculation in Appendix~\ref{sec:app}.

As stressed above, in this section we focus only on $\calP_{(1)}$ from the bulk interaction only at $\tau_\ue$ and
do not discuss whether other one-loop terms yield non-negligible contributions.
Provided that this gives the dominant contribution, the perturbativity requires   
\bae{
	\label{perreq} \f{\calP_{(1)}(k_*)}{\calP_\SR(k_*)} = \f{1}{4} (\Delta\eta)^2 \calP_\SR(k_*) \mk{\f{k_\ue}{k_*}}^{-0.03} \mk{\f{k_\ue}{k_\us}}^{0.03} I\mk{\f{k_\ue}{k_\us}} \ll 1.
}
Here, $\calP_\SR=2.1\times 10^{-9}$ and $\Delta\eta =-2\beta$.
We adopt $k_\ue$ as the \ac{PBH} scale since it corresponds to the wavenumber for the peak of the power spectrum in the transient constant roll scenario as shown in Fig.~\ref{fig: calPCR}.
We consider two cases: the case of the \ac{PBH} as dark matter with $k_\ue=\mO(10^{14})\,\si{Mpc^{-1}}$, and the LIGO--Virgo--KAGRA black holes with $k_\ue=\mO(10^{6})\,\si{Mpc^{-1}}$. 
For a given $k_\ue$, the perturbativity requirement~\eqref{perreq} yields an upper bound on $k_\ue/k_\us$:
\beq \label{lc} \f{k_\ue}{k_\us} \ll \ell_\crit. \eeq

We numerically evaluate the critical value $\ell_\crit$ for specific values of $\beta$ and $k_\ue$ as follows.
First, we substitute the analytic approximation~\eqref{integral-app} of the integral into \eqref{perreq} neglecting $\ln k_\ue/k_\us$ in \eqref{integral-app} and $\mk{k_\ue/k_*}^{-0.03}$ in \eqref{perreq} and analytically derive an approximated form of the critical value.
Using this value as an initial guess, we perform a numerical root-finding algorithm with the analytic approximation~\eqref{integral-app} without neglecting $\ln k_\ue/k_\us$ and $\mk{k_\ue/k_*}^{-0.03}$.
Finally, using the root obtained in the second step as an initial guess, we perform a numerical root-finding algorithm with the integral $I$ numerically evaluated without approximation and obtain the critical value $\ell_\crit$.
The fractional error between the root obtained in the second step and the critical value $\ell_\crit$ remains $\mO(10^{-2})$ for $-3\leq \beta \lesssim -0.7$ but reaches $\mO(10^{-1})$ for $\beta \geq -0.7$ so the final step is important.

\begin{figure}[t]
\centering
\includegraphics[width=.48\textwidth]{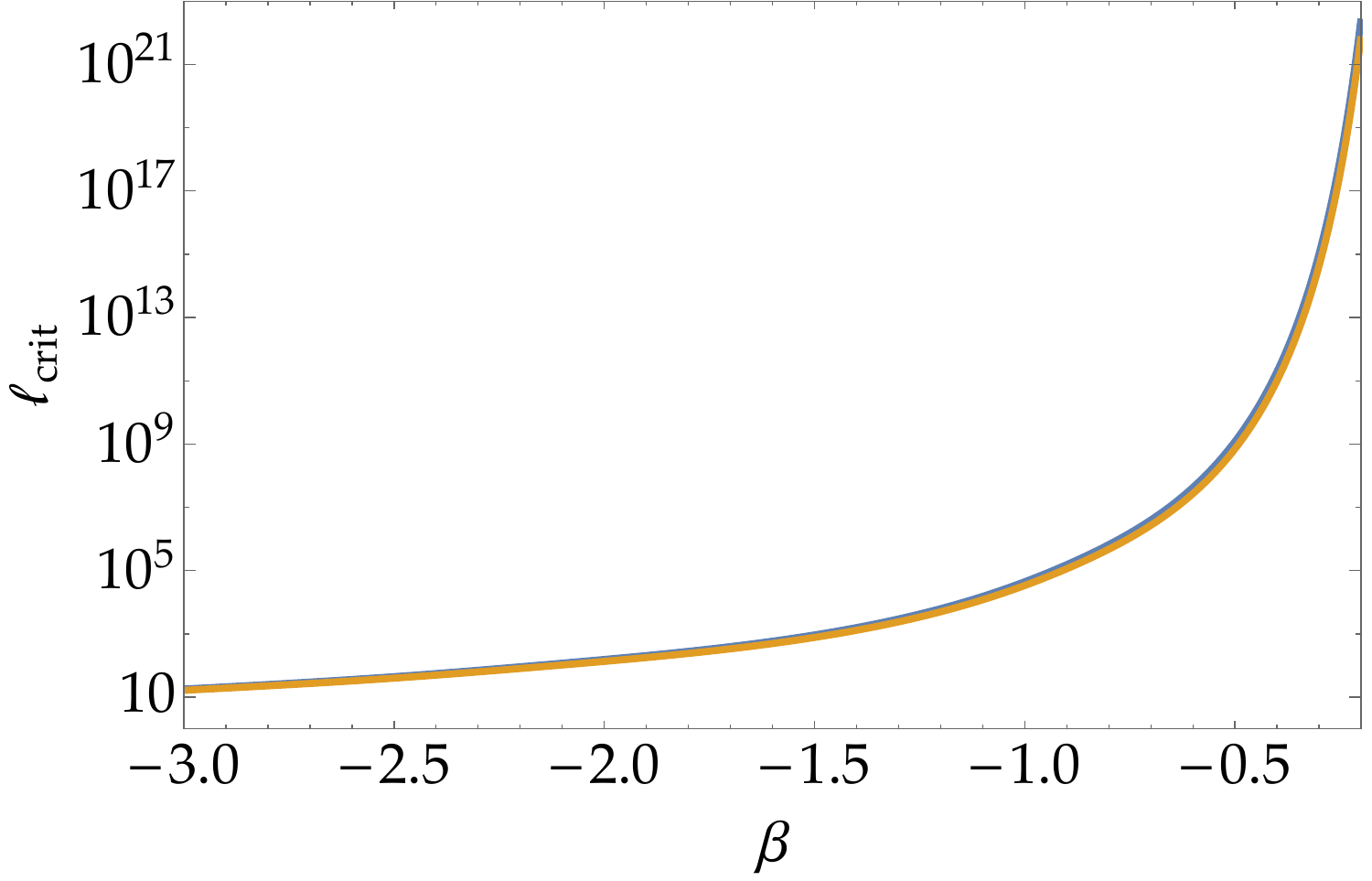}
\includegraphics[width=.48\textwidth]{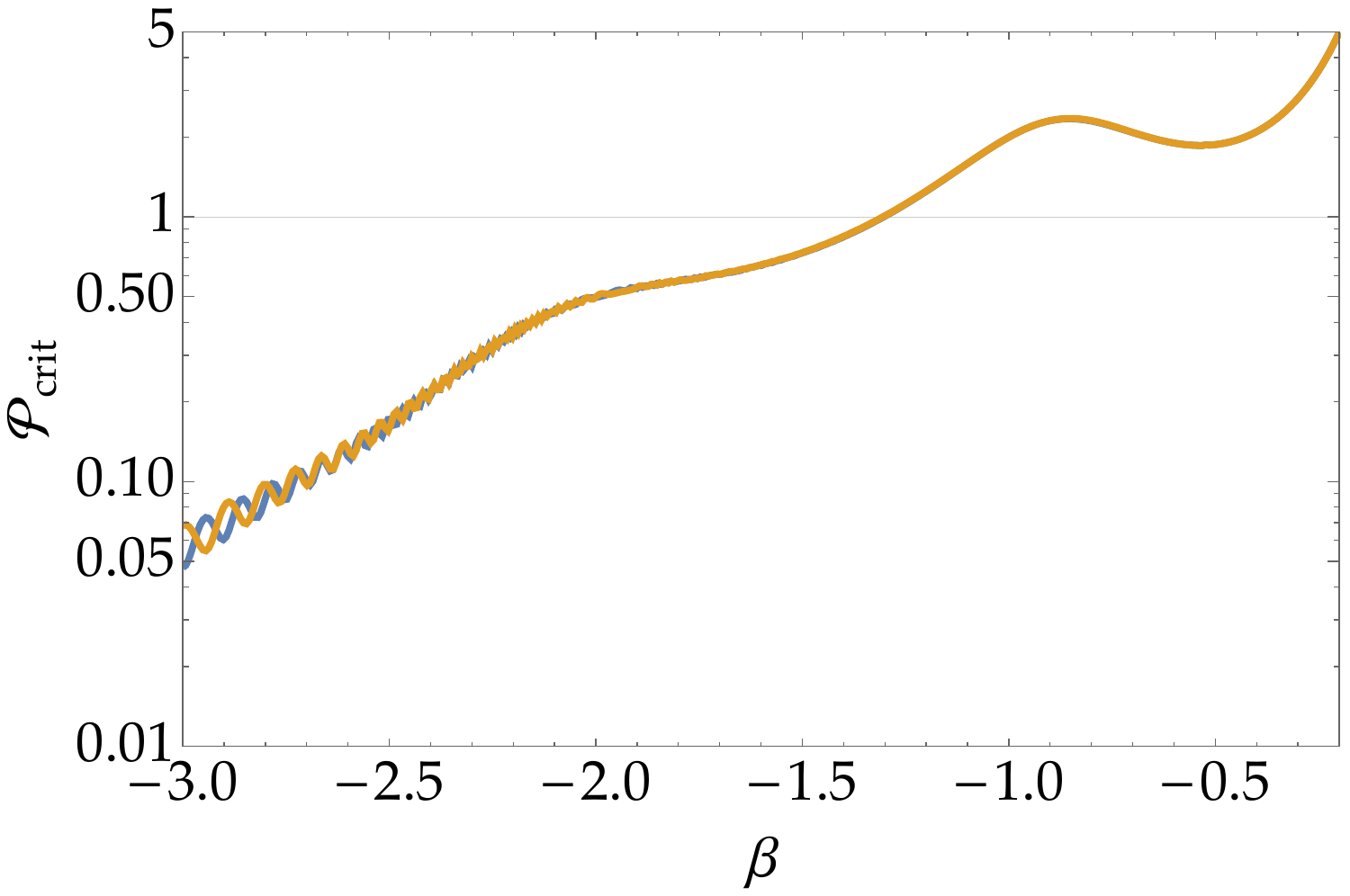}
\caption{The critical values $\ell_\crit$ and $\calP_\crit$ from the perturbativity requirement (see Eqs.~\eqref{lc} and \eqref{D2c}) for $k_\ue=10^{14}\,\si{Mpc^{-1}}$ (blue) and $10^6\,\si{Mpc^{-1}}$ (orange). 
}
\label{fig:lcD2c}
\end{figure}

In the left panel of Fig.~\ref{fig:lcD2c}, we represent the critical value $\ell_\crit$ as a function of $\beta$ for the two cases with $k_\ue=10^{14}\,\si{Mpc^{-1}}$ and $10^6\,\si{Mpc^{-1}}$.
For the ultra slow-roll case with $\beta=-3$, we obtain $\ell_\crit\simeq 18$ and $17$ for the two cases.
In contrast, larger values are not prohibited for $\ell_\crit$ for general constant-roll models.
It implies that a larger wavenumber range of the amplification of the power spectrum is compatible with the perturbativity requirement. 
We also see that the critical value $\ell_\crit$ is not sensitive to $k_\ue$. 

The upper bound of $k_\ue/k_\us$ puts a constraint on the power spectrum on PBH scales as 
\bae{
	\label{D2c} \calP_{\rm PBH} \ll \calP_\crit,
}
where from Eq.~\eqref{PCR} the critical value for the power spectrum is given by
\bae{
	\calP_\crit \equiv \eval{\mk{\f{k_\us}{k_*}}^{-0.03} \calP_\SR(k_*) \abs{F(k_\ue,\tau_\ue)}^2}_{k_\ue/k_\us=\ell_\crit}.
}
In the right panel of Fig.~\ref{fig:lcD2c}, we depict the critical value $\calP_\crit$ for $k_\ue=10^{14}\,\si{Mpc^{-1}}$ and $10^6\,\si{Mpc^{-1}}$.
While there are small oscillations with respect to the values of $\beta$, the critical value $\calP_\crit$ is not so sensitive to $k_\ue$. 
As expected, while $\calP_\crit\sim \mO(10^{-2})$ for the transient ultra slow-roll scenario with $\beta=-3$, larger values are allowed for the transient constant-roll scenario. 
In particular, for the parameter range $-3/2<\beta<0$, where the constant-roll inflation is an attractor, $\calP_\crit \gtrsim \mO(1)$.
Note that \acp{PBH} can be formed when $\calP_{\rm PBH} \gtrsim \mO(10^{-2})$, which is roughly two orders of magnitude below the critical value $\calP_\crit$.
Hence, the perturbativity requirement on the one-loop contribution $\calP_{(1)}(k_*)$ does not rule out the \ac{PBH} production in the transient constant-roll inflation.

\section{Conclusions}
\label{sec:conc}

In canonical single-field inflation, the \ac{PBH} production requires a transient violation of slow-roll.  The representative scenario that satisfies the condition is the transient ultra slow-roll inflation, which has been extensively explored recently.  More generally, we can consider the transient constant-roll inflation, where the power spectrum can be enhanced in non-/attractor inflationary dynamics.  In the present paper, we investigated the squeezed bispectrum in the transient constant-roll inflation and also demonstrated how the one-loop corrections are modified from the transient ultra slow-roll inflation, picking up the representative terms originating from $\dot\eta$ term in the cubic action, as the first step to the full investigation of the one-loop analysis.

Applying the Feynman rule in the Schwinger--Keldysh formalism, we calculated the squeezed bispectrum both after and during the constant-roll phase.  We confirmed that the Maldacena's consistency relation holds for the bispectrum between the \ac{CMB} scale mode and the \ac{PBH} scale mode and hence the \ac{PBH} distribution is not modulated beyond the adiabatic perturbation on large scales. On the other hand, for the non-attractor model, the bispectrum between the \ac{PBH} scale modes themselves violates Maldacena's consistency relation. Therefore, the detailed prediction of the \ac{PBH} abundance, its clustering behaviour on small scales, or the corresponding induced \ac{GW} requires special care to take into account the effects of the violation of the consistency relation.

We investigated the one-loop corrections originating from the $\dot\eta$ term in the cubic action in the transient constant-roll inflation and addressed whether the perturbativity requirement on those terms rules out the \ac{PBH} production.
We clarified that, compared to the transient ultra slow-roll inflation, a larger wavenumber range is allowed for the amplification of the power spectrum, and hence a larger amplification is allowed. 
As a result, we found that the \ac{PBH} production is not ruled out from the perturbativity requirement, for both mass scales of the \ac{PBH} as dark matter or LIGO--Virgo--KAGRA black holes.

Recently, the one-loop corrections in the \ac{PBH} production from the transient ultra slow-roll inflation~\cite{Kristiano:2022maq,Kristiano:2023scm,Riotto:2023hoz,Riotto:2023gpm,Choudhury:2023vuj,Choudhury:2023jlt,Choudhury:2023rks,Firouzjahi:2023aum,Firouzjahi:2023ahg} and the resonance model~\cite{Inomata:2022yte} have been extensively explored, and it has been actively debated whether the perturbativity requirement rules out the \ac{PBH} production in canonical single-field inflation.
Our results show that the \ac{PBH} production from the transient constant-roll inflation is not prohibited from the perturbativity requirement at least on the one-loop corrections originating from $\dot\eta$ term in the cubic action.\footnote{Refs.~\cite{Firouzjahi:2023aum,Firouzjahi:2023ahg} show that smooth transitions between slow-roll and ultra slow-roll phases can reduce the loop correction as another counterexample.}
However, we reiterate that in the present paper, we focused on particular terms of the one-loop corrections, which are often regarded as a dominant term in the literature.
Other one-loop terms may or may not contribute in the transient constant-roll inflation.
It requires a careful study to take into account all the relevant one-loop terms, which we leave for future work.

\acknowledgments
We thank Ryo Saito, Takahiro Terada, and Junsei Tokuda for their helpful discussions.
This work was supported by Japan Society for the Promotion of Science (JSPS) Grants-in-Aid for Scientific Research (KAKENHI) Grant No.~JP22K03639 (H.M.) and No.~JP21K13918 (Y.T.).

\appendix

\section{Reconstruction of the step-function-like transition}\label{sec: appendix}

In this appendix, we explicitly reconstruct a potential that realizes the step-function-like transition, which we consider in the main text:
\bae{\label{eq: eta}
	\eta=\partial_N\ln\epsilon=\bce{
		0, & \tau<\tau_\us, \\
		2\beta, & \tau_\us\leq\tau<\tau_\ue, \\
		0, & \tau_\ue\leq\tau,
	}
}
with transition times $\tau_\us<\tau_\ue$, the constant-roll parameter $\beta$, and the first slow-roll parameter $\epsilon=-\partial_N\ln H$.
First, it is solved as
\bae{
	&\ln\epsilon(N)=\int_0^N\eta\dd{N}+\ln\epsilon_\ui=\bce{
		\ln\epsilon_\ui, & N<N_\us, \\
		2\beta(N-N_\us)+\ln\epsilon_\ui, & N_\us\leq N<N_\ue, \\
		2\beta(N_\ue-N_\us)+\ln\epsilon_\ui, & N_\ue\leq N,
	} \nonumber \\
	\Rightarrow{} & \epsilon(N)=\bce{
		\epsilon_\ui, & N<N_\us, \\
		\epsilon_\ui\ee^{2\beta(N-N_\us)}, & N_\us\leq N<N_\ue, \\
		\epsilon_\ui\ee^{2\beta(N_\ue-N_\us)}, & N_\ue\leq N,
	}
}
where $\epsilon_\ui$ is the initial value of $\epsilon$ at the initial time which we choose $N=0$ without loss of generality.
$N_\us$ and $N_\ue$ are e-folding times corresponding to $\tau_\us$ and $\tau_\ue$, respectively.
As $\epsilon$ is also related to the inflaton's velocity by $\epsilon=\frac{1}{2}\pqty{\dv{\phi}{N}}^2$, if one assumes $\dv*{\phi}{N}>0$ without loss of generality, it is solved as
\bae{
	&\dv{\phi}{N}=\sqrt{2\epsilon}=\bce{
		\sqrt{2\epsilon_\ui}, & N<N_\us, \\
		\sqrt{2\epsilon_\ui}\ee^{\beta(N-N_\us)}, & N_\us\leq N<N_\ue, \\
		\sqrt{2\epsilon_\ui}\ee^{\beta(N_\ue-N_\us)}, & N_\ue\leq N,
	} \nonumber \\
	\Rightarrow{} & \phi(N)=\bce{
		\sqrt{2\epsilon_\ui}N, & N<N_\us, \\
		\sqrt{2\epsilon_\ui}\bqty{\frac{1}{\beta}\pqty{\ee^{\beta(N-N_\us)}-1}+N_\us}, & N_\us\leq N<N_\ue, \\
		\sqrt{2\epsilon_\ui}\bqty{\ee^{\beta(N_\ue-N_\us)}(N-N_\ue)+\frac{1}{\beta}\pqty{\ee^{\beta(N_\ue-N_\us)}-1}+N_\us}, & N_\ue\leq N,
	}
}
where we set the initial field value $\phi(N=0)=0$ without loss of generality. Its inverse function is found as
\bae{\label{eq: N of phi}
	N(\phi)=\bce{
		\frac{\phi}{\sqrt{2\epsilon_\ui}}, & \phi<\phi_\us, \\
		N_\us+\frac{1}{\beta}\ln\bqty{\beta\pqty{\frac{\phi}{\sqrt{2\epsilon_\ui}}-N_\us}+1}, & \phi_\us\leq\phi<\phi_\ue, \\
		N_\ue+\ee^{-\beta(N_\ue-N_\us)}\bqty{\frac{\phi}{\sqrt{2\epsilon_\ui}}-\frac{1}{\beta}\pqty{\ee^{\beta(N_\ue-N_\us)}-1}-N_\us}, & \phi_\ue\leq\phi,
	}
}
where $\phi_\us=\phi(N_\us)$ and $\phi_\ue=\phi(N_\ue)$.
$\epsilon$ is further related to the Hubble parameter by $\epsilon=-\partial_N\ln H$, which means
\bae{
	&\ln H(N)=\bce{
		\ln H_\ui-\epsilon_\ui N, & N<N_\us, \\
		\ln H_\ui-\frac{\epsilon_\ui}{2\beta}\pqty{\ee^{2\beta(N-N_\us)}-1}-\epsilon_\ui N_\us, & N_\us\leq N<N_\ue, \\
		\ln H_\ui-\epsilon_\ui\ee^{2\beta(N_\ue-N_\us)}(N-N_\ue)-\frac{\epsilon_\ui}{2\beta}\pqty{\ee^{2\beta(N_\ue-N_\us)}-1}-\epsilon_\ui N_\us, & N_\ue\leq N,
	} \nonumber \\
	\Rightarrow{} & H(N)=\bce{
		H_\ui\ee^{-\epsilon_\ui N}, & N<N_\us, \\
		H_\ui\exp\bqty{-\frac{\epsilon_\ui}{2\beta}\pqty{\ee^{2\beta(N-N_\us)}-1}-\epsilon_\ui N_\us}, & N_\us\leq N<N_\ue, \\
		H_\ui\exp\bqty{-\epsilon_\ui\ee^{2\beta(N_\ue-N_\us)}(N-N_\ue)-\frac{\epsilon_\ui}{2\beta}\pqty{\ee^{2\beta(N_\ue-N_\us)}-1}-\epsilon_\ui N_\us}, & N_\ue\leq N,
	}
}
where $H_\ui$ is the initial value of the Hubble parameter. Combining it with Eq.~\eqref{eq: N of phi}, one finds $H$ as a function of $\phi$ as
\bae{
	H(\phi)=\bce{
		H_\ui\ee^{-\sqrt{\frac{\epsilon_\ui}{2}}\phi}, & \phi<\phi_\us, \\
		H_\ui\exp\bqty{-\frac{\epsilon_\ui\beta}{2}\pqty{\frac{\phi}{\sqrt{2\epsilon_\ui}}-N_\us}^2-\sqrt{\frac{\epsilon_\ui}{2}}\phi}, & \phi_\us\leq\phi<\phi_\ue, \\[10pt]
		\bmbe{\textstyle H_\ui\exp\left[-\epsilon_\ui\ee^{\beta(N_\ue-N_\us)}\pqty{\frac{\phi}{\sqrt{2\epsilon_\ui}}-\frac{1}{\beta}\pqty{\ee^{\beta(N_\ue-N_\us)}-1}-N_\us} \right. \\[-10pt]
		\textstyle \left. -\frac{\epsilon_\ui}{2\beta}\pqty{\ee^{2\beta(N_\ue-N_\us)}-1}-\epsilon_\ui N_\us \right],} & \phi_\ue\leq\phi.
	}
}
The Hamilton--Jacobi equation finally connects the Hubble parameter to the potential by
\bae{\label{eq: reconst pot}
	V(\phi)&=3H^2(\phi)-2{H'(\phi)}^2 \nonumber \\
	&=\bce{
		(3-\epsilon_\ui)H_\ui^2\ee^{-\sqrt{2\epsilon_\ui}\phi}, & \phi<\phi_\us, \\[5pt]
		\bmbe{\textstyle -\frac{1}{2} H_\ui^2 \left[\beta^2 \phi^2-2 \sqrt{2\epsilon_\ui} \beta \phi (\beta N_\us-1)+2 \epsilon_\ui (\beta N_\us-1)^2-6\right] \\[-10pt]
		\textstyle\times\exp \left[-\frac{1}{2} \phi \left(\beta\phi +2 \sqrt{2\epsilon_\ui}\right)-\beta N_\us^2 \epsilon_\ui+\sqrt{2\epsilon_\ui} \beta N_\us \phi \right],} & \phi_\us\leq\phi<\phi_\ue, \\[5pt]
		\bmbe{\textstyle \left(3-\epsilon_\ui \ee^{2 \beta (N_\ue-N_\us)}\right)H_\ui^2 \\[-10pt]
		\textstyle \times\exp\left(\frac{\epsilon_\ui \left(\ee^{\beta (N_\ue-N_\us)}-1\right) \left(\ee^{\beta (N_\ue-N_\us)}+2 \beta N_\us-1\right)}{\beta}-\sqrt{2\epsilon_\ui} \phi \ee^{\beta (N_\ue-N_\us)}\right),} & \phi_\ue\leq\phi.
	}
}
For this potential, one can numerically solve the inflaton's dynamics through its equation of motion
\bae{
	\ddot{\phi}+3H\dot{\phi}+V'(\phi)=0 \qc 3H^2=\frac{\dot{\phi}^2}{2}+V(\phi),
}
and reproduce the step-function-like transition~\eqref{eq: eta} as shown in Fig.~\ref{fig: SR params}.

\bfe{width=0.7\hsize}{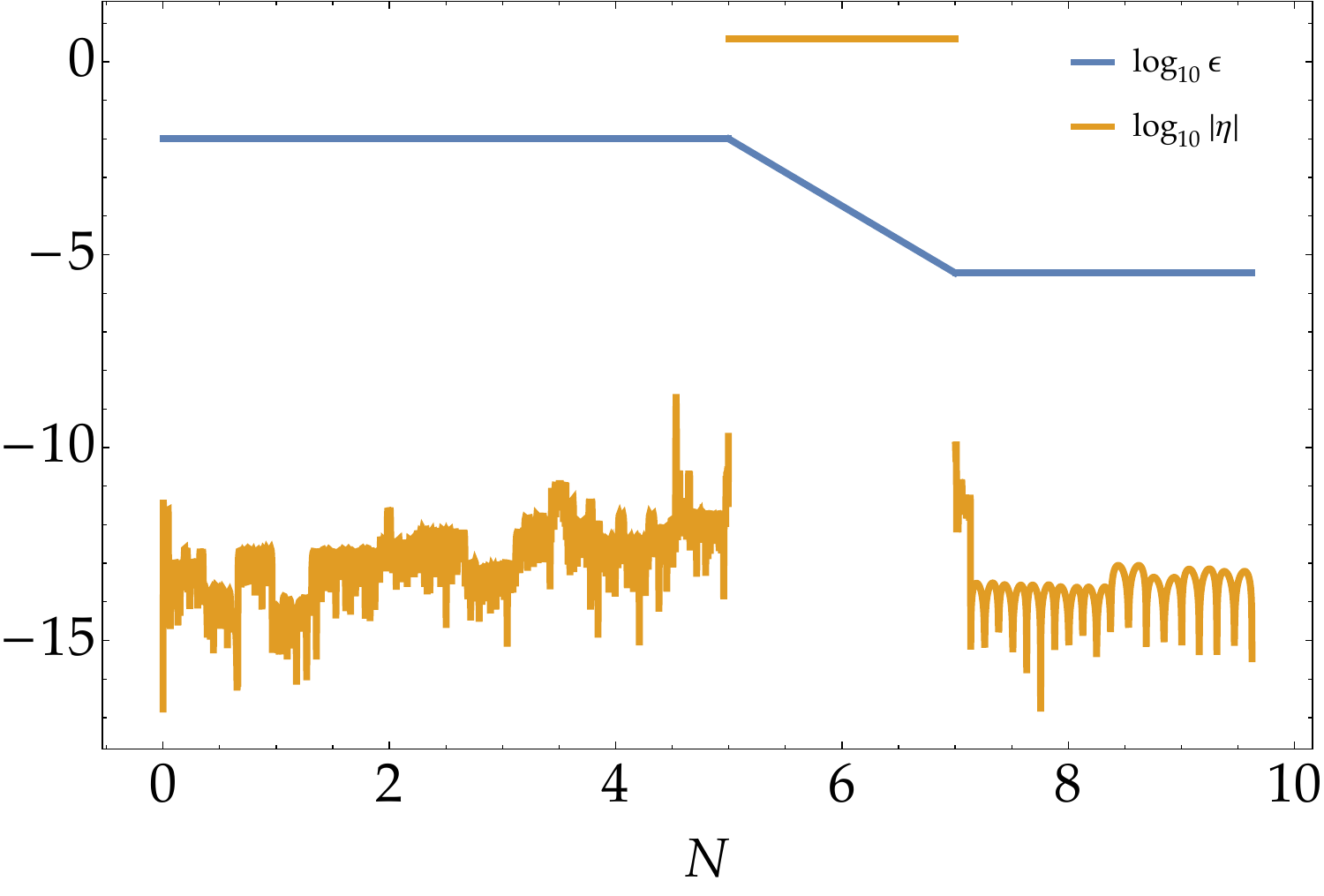}{Numerical results of the slow-roll parameters $\log_{10}\epsilon$ (blue) and $\log_{10}\abs{\eta}$ (orange) for the inflarinary dynamics with the reconstructed potential~\eqref{eq: reconst pot} with the parameter set $\beta=-2$, $\epsilon_\ui=10^{-2}$, $H_\ui=10^{-5}$, $N_\us=5$, and $N_\ue=7$. The noisy feature of $\eta$ would be caused by the numerical error.}{fig: SR params}

\section{\texorpdfstring{Evaluation of the integral~\eqref{integral-def}}{Evaluation of the integral (4.9)}}
\label{sec:app}

In this appendix, we analytically evaluate the integral~\eqref{integral-def} under the assumption $\ell_\ue\equiv k_\ue/k_\us\gg 1$ and derive the approximated form~\eqref{integral-app}.
For $k\simeq k_\ue$, the function $F(k_\ue,\tau_\ue)$ in the integrand scales as $|F(k_\ue,\tau_\ue)|^2 \simeq \ell_\ue^{-2\beta} (\beta^4 + 6 \beta^3 + 13 \beta^2 + 12 \beta + 8) / 4$ for $\ell_\ue \gg 1$.
Below we shall consider an approximation for $k_\us \leq k \leq k_\ue$.

First, we decompose the integrand into the nonoscillating part $G_1(\ell)$ and oscillating part $G_2(\ell)$:
\bae{
    \f{|F(\ell k_\us,\tau_\ue)|^2}{\ell} = G_1(\ell) + G_2(\ell) + G_2^*(\ell),
}
where
\beae{
    &G_1(\ell)=\bmte{\frac{\ell_\ue^{-2 (\beta +1)} }{2 \ell^3[\ell^2 + b(b-1) ]^2} (\ell^2 + b^2 \ell_\ue^2) \\ 
    \times [ 2 \ell^6 + \left(\beta^2+2 b(b-1) \right) \ell^4 + \left(b^2 (\beta -1)^2+(\beta +1)^2-2 b\right) \ell^2 + b^2 \beta ^2 ],} \\
    &G_2(\ell) = \bmte{\frac{i \ell_\ue^{-2 (\beta +1)} \ee^{-2 i \ell (1-1/\ell_\ue)} }{4 \ell^3 [\ell^2+b(b-1)]^2} (\ell+i b \ell_\ue)^2 
    [\beta \ell^2 - i (b (\beta -1)+\beta +1) \ell - b \beta ] \\
    \times [2 \ell^3 + i (-2 b+\beta +2) \ell^2 + (b-1) (\beta +1) \ell  + i b \beta ].}
}
Since $|b(b-1)| \leq 1/4$ for $-3<\beta<0$, we approximate $\ell^2 + b(b-1) \simeq \ell^2$ in the denominator of $G_1$ and $G_2$.

For the nonoscillating part, we simply integrate it, pick up terms with the highest power of $\ell_\ue$, and arrive at 
\begin{align}
I_1 &\equiv \int_1^{\ell_\ue} G_1(\ell) \dd{\ell} \notag\\
&\bmbe{\simeq \ell_\ue^{-2 \beta } \left[ \frac{1}{4} (\beta +1)^2 (\beta +2)^2 \ln \ell_\ue + \frac{1}{2} \right. \\
\left.+ \frac{1}{384} \beta (\beta +1)^2 (\beta +2)^2 \left(5 \beta ^5+24 \beta ^4+44 \beta ^3+72
\beta ^2+143 \beta +72\right)\right] .}
\end{align}
The nonoscillating part yields the dominant contribution.

For the oscillating part, the contribution to the integral mainly comes from $\ell\sim 1$ since $G_2$ is rapidly oscillating with a slowly varying envelope, and hence the contribution to the integral is negligible for $\ell\gg 1$.
Hence, we Taylor expand the integrand around $\ell=1$ up to the first order, integrate it, and then pick up terms with the highest power of $\ell_\ue$.  
As a result, we obtain 
\begin{align}
    I_2 &\equiv \int_1^{\ell_\ue} [G_2(\ell)+G_2^*(\ell)] \dd{\ell} \notag\\
    &\bmbe{\simeq \frac{\ell_\ue^{-2 \beta }}{960} \beta  (\beta +1)^2 (\beta +2)^2 \left[ 4 \left(7 \beta ^4+24 \beta ^3+27 \beta^2+26 \beta -4\right) \sin 2 \right. 
    \\
    \left. - \left(12 \beta ^5+58 \beta ^4+65 \beta ^3+48 \beta ^2+101\beta +36\right) \cos 2 \right] .}
\end{align}

\begin{figure}[t]
\centering
\includegraphics[width=0.7\hsize]{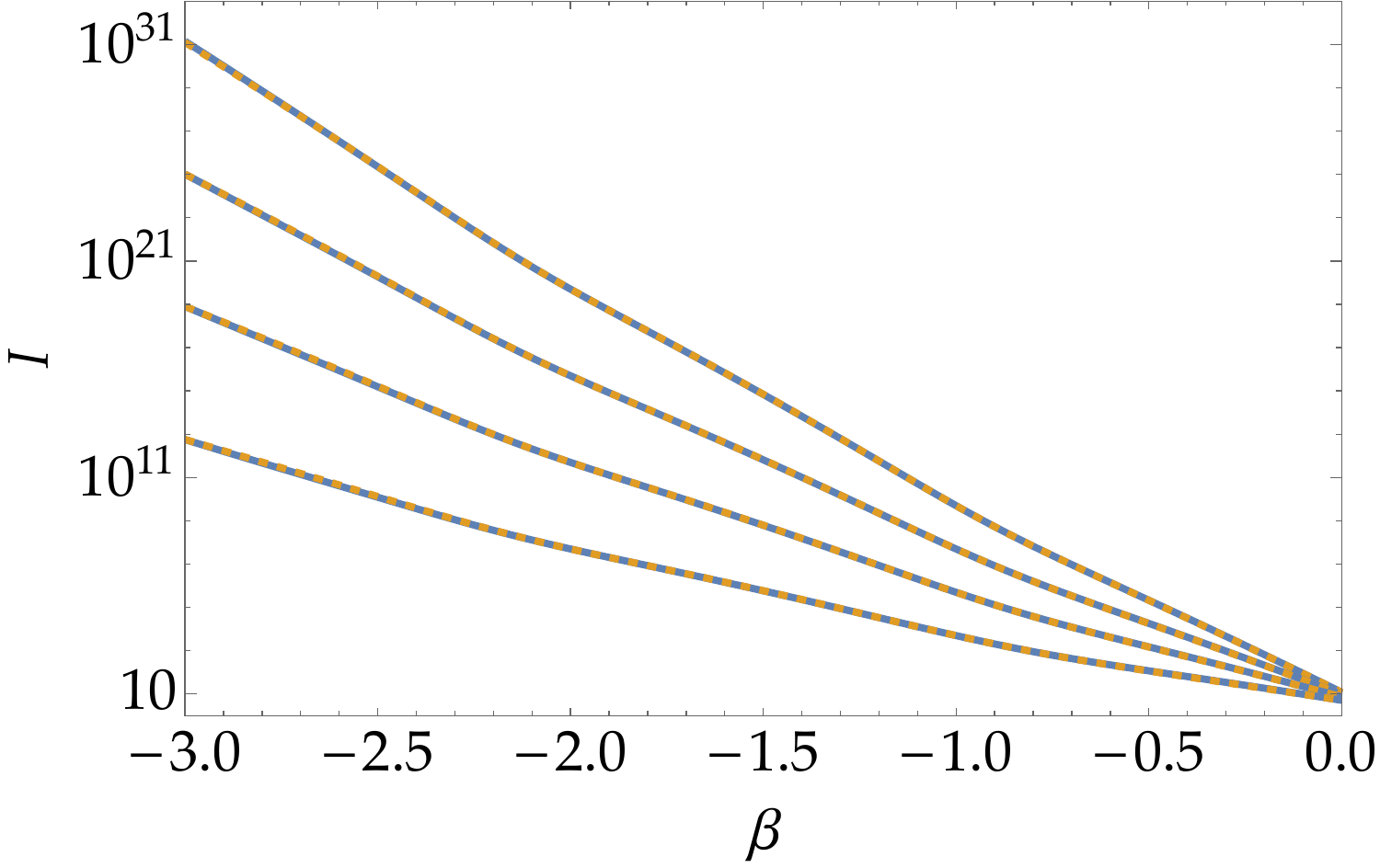}
\caption{
The integral $I$ as a function of the constant roll parameter $\beta$. Numerical integration of the exact form \eqref{integral-def} (blue, solid) and the approximation \eqref{integral-app} (orange, dashed) are shown for $\ell_\ue\equiv k_\ue/k_\us = 10^2$, $10^3$, $10^4$, and $10^5$ (from bottom to top). }
\label{fig:integral}
\end{figure}

With these results, we obtain the analytic approximation of the integral $I=I_1+I_2$ as in \eqref{integral-app}.
One can recover the result $\mk{\f{k_\ue}{k_\us}}^6 \mk{1.1+\ln \f{k_\ue}{k_\us}}$ obtained in Ref.~\cite{Kristiano:2022maq} by substituting $\beta=-3$.
In Fig.~\ref{fig:integral}, we compare the analytic formula and numerical calculation for the parameter range $-3<\beta<0$, and confirm that they are in good agreement.

\bibliographystyle{JHEP}
\bibliography{refs}
\end{document}